\documentclass[twocolumn,notitlepage,prl,superscriptaddress,longtable,longbibliography]{revtex4-1}
\usepackage{multirow}
\usepackage{graphicx}  
\usepackage{dcolumn}   
\usepackage{tabularx}
\usepackage{tabu}
\usepackage{bm}        
\usepackage{amssymb}  
\usepackage{amsmath}
\usepackage{mathrsfs}
\usepackage{eqnarray,amsmath}
\usepackage{mhchem}
\usepackage{amsthm}
\usepackage{amsmath}
\usepackage{amssymb}
\usepackage{mathdots}
\usepackage{graphicx}
\usepackage{mathrsfs}
\usepackage{longtable}
\usepackage[colorinlistoftodos]{todonotes}
\usepackage[colorlinks=true, allcolors=blue]{hyperref}
\usepackage{physics}
\usepackage{float}
\usepackage{hyperref}
\usepackage[english,french]{babel}

\providecommand{\tabularnewline}{\\}

\usepackage{braket}
\usepackage{hyperref}
\usepackage{colortbl}
\usepackage[colorinlistoftodos]{todonotes}

\makeatother

\usepackage{babel}

\begin{document}
\global\long\def\sx{\sigma_{x}}
\global\long\def\sy{\sigma_{y}}
\global\long\def\sz{\sigma_{z}}
\global\long\def\tx{\tau_{x}}
\global\long\def\ty{\tau_{y}}
\global\long\def\tz{\tau_{z}}
\global\long\def\snx{\sin k_{x}}
\global\long\def\sny{\sin k_{y}}
\global\long\def\csx{\cos k_{x}}
\global\long\def\csxx{\cos \frac{k_{x}}{2}}
\global\long\def\snxx{\sin \frac{k_{x}}{2}}
\global\long\def\csyy{\cos \frac{\sqrt{3}k_{y}}{2}}
\global\long\def\snyy{\sin \frac{\sqrt{3}k_{y}}{2}}
\global\long\def\snyyt{\sin \sqrt{3}k_{y}}
\global\long\def\csyyt{\cos \sqrt{3}k_{y}}
\global\long\def\csy{\cos k_{y}}
\global\long\def\px{p_{x}}
\global\long\def\py{p_{y}}
\global\long\def\pz{p_{z}}
\global\long\def\up{\uparrow}
\global\long\def\dn{\downarrow}
\title{Phonon-induced Floquet second-order topological phases protected by space-time symmetries}

\author{Swati Chaudhary}
\email{swatich@caltech.edu}
\affiliation{Institute of Quantum Information and Matter and Department of Physics, California Institute of Technology, Pasadena, CA 91125, USA}
\author{Arbel Haim}
\email{arbelh@caltech.edu}
\affiliation{Institute of Quantum Information and Matter and Department of Physics, California Institute of Technology, Pasadena, CA 91125, USA}
\affiliation{Walter Burke Institute for Theoretical Physics, California Institute of Technology, Pasadena, CA 91125, USA}
\author{Yang Peng}
\email{yangpeng@caltech.edu}
\affiliation{Institute of Quantum Information and Matter and Department of Physics, California Institute of Technology, Pasadena, CA 91125, USA}
\affiliation{Walter Burke Institute for Theoretical Physics, California Institute of Technology, Pasadena, CA 91125, USA}
\affiliation{Department of Physics and Astronomy, California State University, Northridge, California 91330, USA}
\author{Gil Refael}
\affiliation{Institute of Quantum Information and Matter and Department of Physics, California Institute of Technology, Pasadena, CA 91125, USA}

\begin{abstract}
The co-existence of spatial and non-spatial symmetries together with appropriate commutation/anticommutation relations between them can give rise to static higher-order topological phases, which host gapless boundary modes of co-dimension higher than one. Alternatively, space-time symmetries in a Floquet system can also lead to anomalous Floquet boundary modes of higher co-dimensions, presumably with alterations in the commutation/anticommutation relations with respect to non-spatial symmetries. We show how a coherently excited phonon mode can be used to promote a spatial symmetry with which the static system is always trivial, to a space-time symmetry which supports non-trivial Floquet higher-order topological phase. We present two examples -- one in class D and another in class AIII where a coherently excited phonon mode promotes the reflection symmetry to a time-glide symmetry such that the commutation/anticommutation relations between spatial and non-spatial symmetries are modified. These altered relations allow the previously trivial system to host gapless modes of co-dimension two at reflection-symmetric boundaries.
\end{abstract}

\maketitle

{\em Introduction.---} 
The topology of electronic band structures of crystals is largely restricted by the existing symmetries
~\cite{Schnyder2008, Kitaev2009, Ryu2010, Turner2010, Hughes2011, Fu2011, Chiu2013, Shiozaki2014}, 
and its nontriviality is reflected in the presence of gapless modes located at the crystal boundaries \cite{Teo2010,
	Chiu2016, Khalaf2018, Khalaf2018prx, Trifunovic2019}. 
For example, in topological insulators \cite{Hasan2010, Qi2011, Bernevig2013book}, where the band topology respects only 
nonspatial symmetries, such as the time-reversal, particle-hole,  and chiral
symmetries, 
the boundary modes are of codimension one (the codimension is the difference between the bulk dimension and the dimension along
which the gapless modes propagate).

Recently, systems \cite{Benalcazar2017, Peng2017, Langbehn2017,
	Benalcazar2017s, Song2017, Schindler2018, Geier2018} 
	were theoretically proposed to support gapless modes
of higher codimensions, because of the additional spatial symmetries coexisting with the nonspatial ones.  The order of
such  higher-order topological insulators is given by the codimension
of the boundary modes.  
On the experimental side, codimension-two boundary modes are observed mostly in metamaterials, 
such as electric circuits \cite{Imhof2018}, photonic \cite{Peterson2018} and phononic \cite{Serra2018, Xue2019, Ni2019} 
systems.
The electronic second-order topological insulator is only realized in Bismuth \cite{Schindler2018natphys}.  

If a spatial symmetry coexists with nonspatial symmetries, the symmetry operator of the former can either commute or 
anticommute with the ones of the latter \cite{Chiu2013, Shiozaki2014}. 
Therefore, the coexistence of a certain spatial symmetry alone is not enough to 
guarantee the possibility of having a nontrivial band topology, but with appropriate commutation or anticommutation
relations between the spatial and nonspatial symmetry operators. 

Very recently, it was demonstrated that in a periodically driven system,  
a new space-time symmetry, such as time-glide or time-screw can emerge,
if the system is invariant under reflection or two-fold rotation,
together with a half-period time translation \cite{Morimoto2017}.
As far as topological classification is concerned, 
such a space-time symmetry can lead to a nontrivial
Floquet band topology, in the same way as its spatial counterpart 
does in a static system, except for a possible alternation 
of the commutation/anticommutation relations with respect to the
nonspatial symmetries \cite{Peng2019, Peng2019b}. 

This result leads to the following interesting question.  
When the commutation/anticommutation relation alternation does occur,  
is it able to periodically drive an initially topological trivial system, 
whose spatial symmetry does not have appropriate relations with respect to the nonspatial symmetries, 
into a nontrivial Floquet higher-order topological insulator? 

In this work, we answser this question by considering phonon-assisted space-time engineering,
which promotes the spatial symmetry (such as reflection) of a static system into a 
space-time symmetry (such as time-glide), without changing the symmetry operator. 
In this way, the relations with respect to the nonspatial symmetries that are 
inappropriate for the spatial symmetry, would become otherwise appropriate for the space-time symmetry. 

{\em Phonon-assisted space-time engineering.---}
One assumption in the electronic band structure of a crystal is
that the lattice is rigid, with ions fixed to their equilibrium positions. 
The success of this assumption in characterizing lots of properties of materials 
is that the energy due to lattice vibrations or phonons is much smaller 
compared to the electronic energy at the equilibrium lattice configuration.  

However, it is known that a coherently excited and macroscopically occupied phonon mode 
can result in ions moving collectively \cite{Kuznetsov1994}.  
When the material is in such a state, the electrons will experience a periodically 
oscillating ionic potential which can no longer be neglected. 
Indeed, such a phonon driven Floquet topological insulator based on graphene has been proposed recently in~Ref.~\cite{Hubener2018}.

It is known that the symmetries of a crystal in equilibrium with a rigid lattice configuration are described
by the space group of the  lattice. The normal modes of the lattice vibrations, namely, the phonons, form
the irreducible representations of this group. To be more specific, 
consider an order-two point group operation $\hat{g}$ which squares to identity; the phonon modes
must have a definite parity under this operation.
Whereas the oscillating potential generated
by the even-parity phonon is invariant under such a point group operation at arbitrary times, 
the one generated by an odd-parity phonon breaks this point group symmetry. 
Nevertheless, the time-dependent potential $V(\boldsymbol{r},t)$ generated by the latter acquires the space-time
symmetry, given by $V(\hat{g}\boldsymbol{r}, t) = V(\boldsymbol{r},t+T/2)$, where $\boldsymbol{r}$ and $t$ 
are the spatial and temporal coordinates, and $T$ is the oscillation period for the phonon. 
Hence, we have managed to promote a spatial order-two symmetry described by $\hat{g}$, 
to a space-time symmetry described by the same operator, 
by coherently exciting a phonon mode that is odd under $\hat{g}$.
This is an example of phonon-assisted space-time engineering.

In this manuscript, we provide two examples in which by promoting a spatial symmetry with operator $\hat{g}$ to a space-time symmetry,
the commutation relations between $\hat{g}$ and nonspatial symmetries become 
appropriate for supporting a nontrivial (Floquet) topological phase, whereas
only a trivial phase exists in a $\hat{g}$-symmetric static system.
In the supplemental material \cite{suppl}, we list all possibilities of realizing a topological
nontrivial Floquet phase from a static trivial system by such phonon-assisted
space-time engineering.

{\em 2D system  in class D/BDI--}
It is known that in the presence of reflection symmetry, class D or BDI exhibit topological behavior characterized by a mirror topological invariant whenever the reflection operator (described by $M$) commutes with the particle-hole operator ($C$) (Table~\ref{table2} in the Supplemental material). We demonstrate that a phonon drive can be used to turn a trivial static system with $\{C,M\}=0$ into non-trivial Floquet topological phase.

Consider a tight-binding model with nearest-neighbor hopping, $t_0$, on a two-dimensional square lattice placed in the proximity to a s-wave superconductor described by the Bloch hamiltonian

\begin{equation}
H_{0}(\mathbf{k})=(m_0-2t_0\csx-2t_0\csy)\tau_{z}+\Delta\tau_{x}+b\sigma_{x}\label{eq:ho_main},
\end{equation}
where $\sigma$ and $\tau$ are Pauli matrices acting on spin and particle-hole degree of freedom respectively. Particle-hole symmetry  is given by $C=\ty\sy$ and reflection is given by $M=\sx$ flipping the coordinates in the $y$ direction. This also has a time-reversal symmetry given by $T=\mathbb{I}$ but it is not relevant for our purpose as  the commutation relation with the time-reversal operator cannot be altered.  Let us consider the effect of a reflection-symmetry breaking phonon which produces a time-dependent Rashba SOC  given by:
\begin{equation}
H(\mathbf{k},t)=2\alpha_0\cos\omega t(\snx\sigma_y-\sny\sigma_x)\tau_z.
\label{classDH}
\end{equation}
An example for  a phonon mode generating such a term is described in the next section.

The full hamiltonian $H_0(\mathbf{k})+H(\mathbf{k},t)$ has a time-glide symmetry with reflection  $M=\sigma_x$, promoted from the static reflection symmetry.
The role of this periodic drive on topological behavior can be understood better by considering the frequency-domain formulation of the Floquet problem and restrict to the two Floquet-zone sector~\cite{Peng2019, Peng2019b}. This two-by-two enlarged Hamiltonian reads

	\begin{equation}
	\begin{split}
	\mathscr{H}(\mathbf{k})&=\begin{pmatrix}H_{0}(\mathbf{k})+\frac{\omega}{2} & H_{1}\\
	H_{\bar{1}} & H_{0}(\mathbf{k})-\frac{\omega}{2}
	\end{pmatrix}
	\end{split}
	\label{eq:Hfl}
	\end{equation}
where $H_n=\frac{1}{T}\int_0^T H(t)e^{-in\omega t}dt$.

\begin{figure}
	
	\includegraphics[scale=0.28]{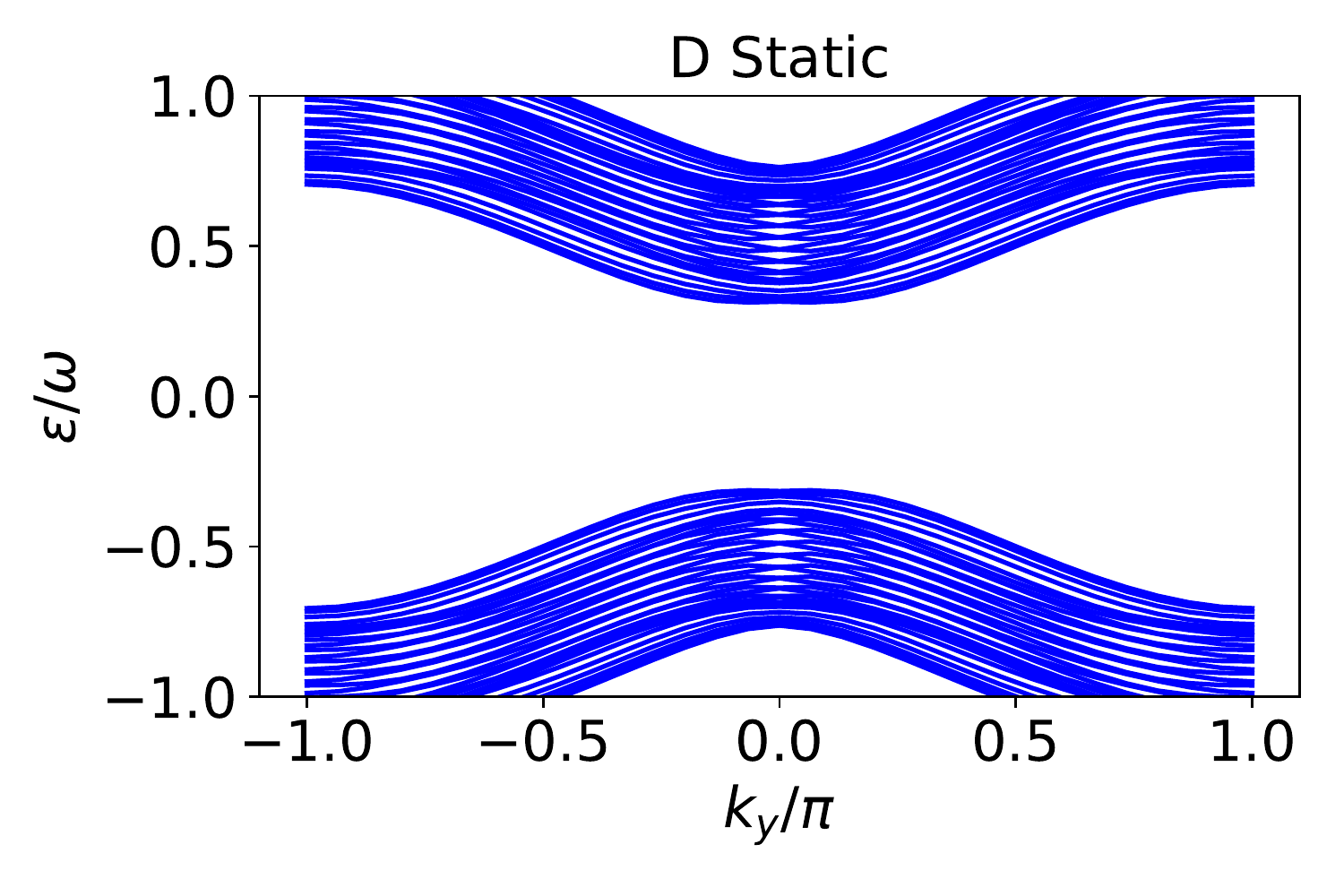}
	\includegraphics[scale=0.28]{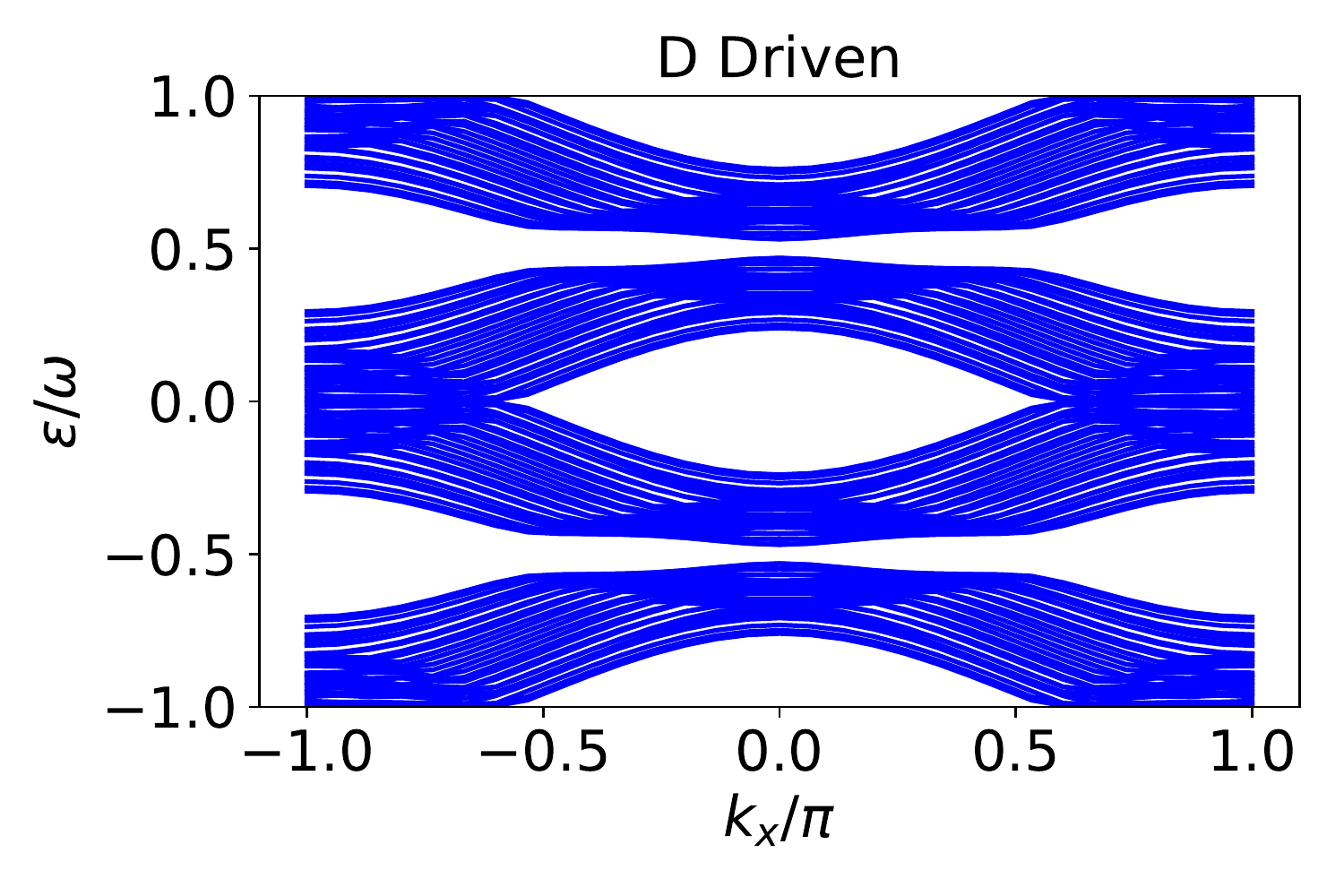}
	
	\includegraphics[scale=0.28]{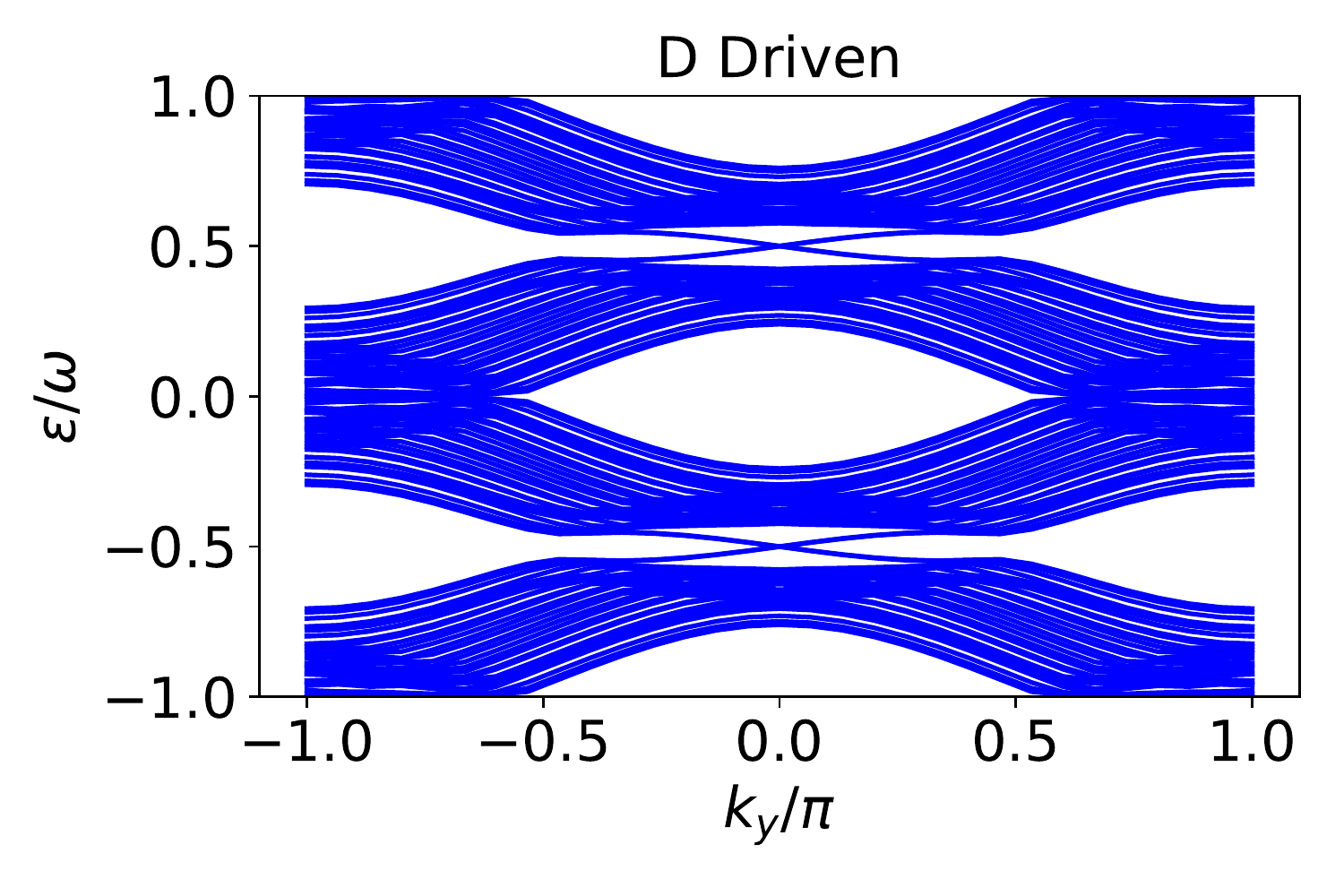}
	\includegraphics[scale=0.28]{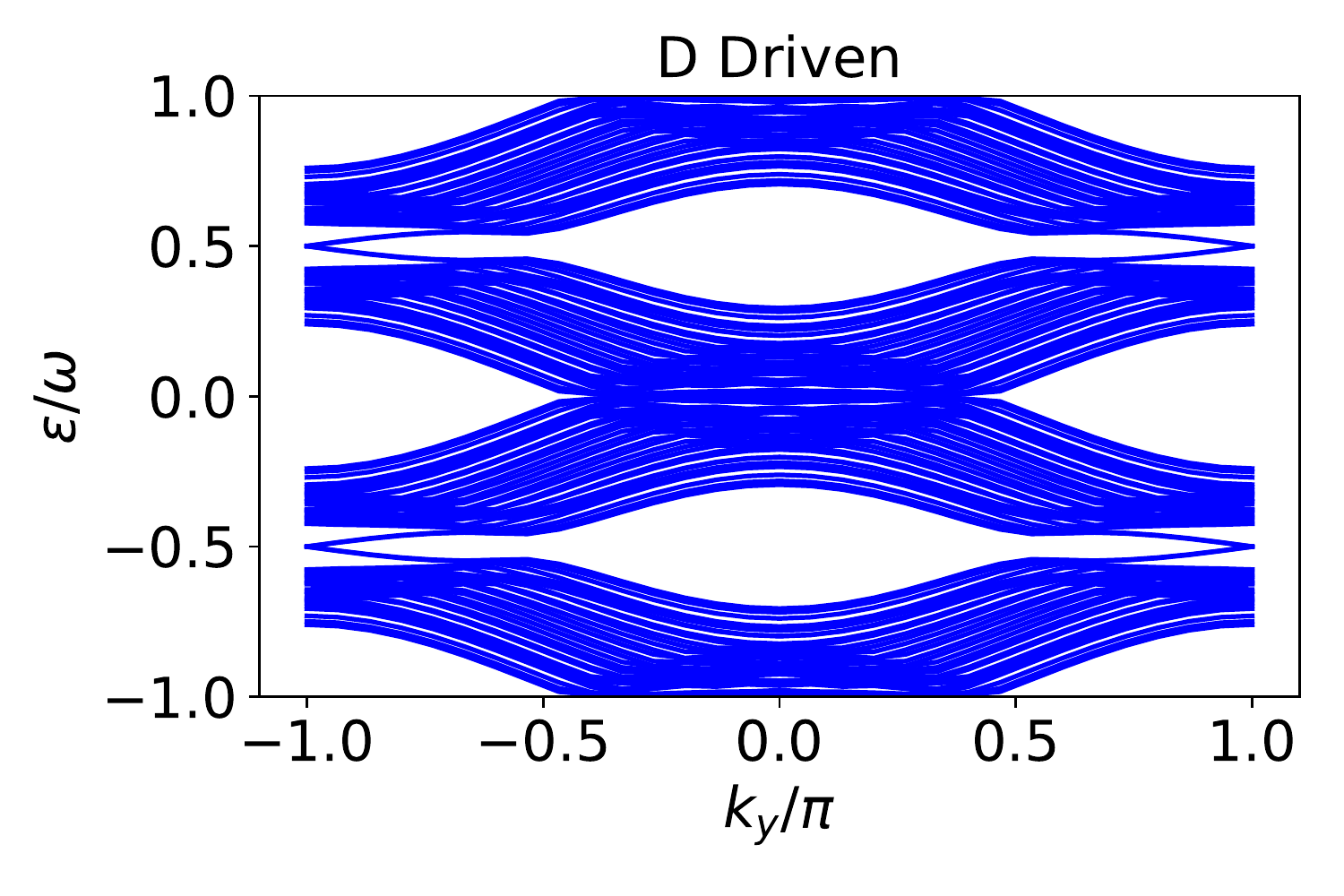}
	
	\caption{Band structure for $m_{0}=\omega/2+1,\ \Delta=0.9,\ \omega=4.8,\ L=15,\ b=0.15,\ \text{and} \ \alpha_0=0.5$ with periodic boundary conditions in one direction.	In the last plot $m_{0}=-\omega/2-1$ and this change of sign results in
		a shift in the position of gapless mode from $k_{y}=0$ to $k_{y}=\pi$ ($\Delta=0.5,\ \alpha_0=1.0$ is shown in Supplemental material). (All energies are in units of $t_0$.)} 
	\label{fig:Band-structure-forfinal}
\end{figure}

\begin{figure}
	\includegraphics[scale=0.27]{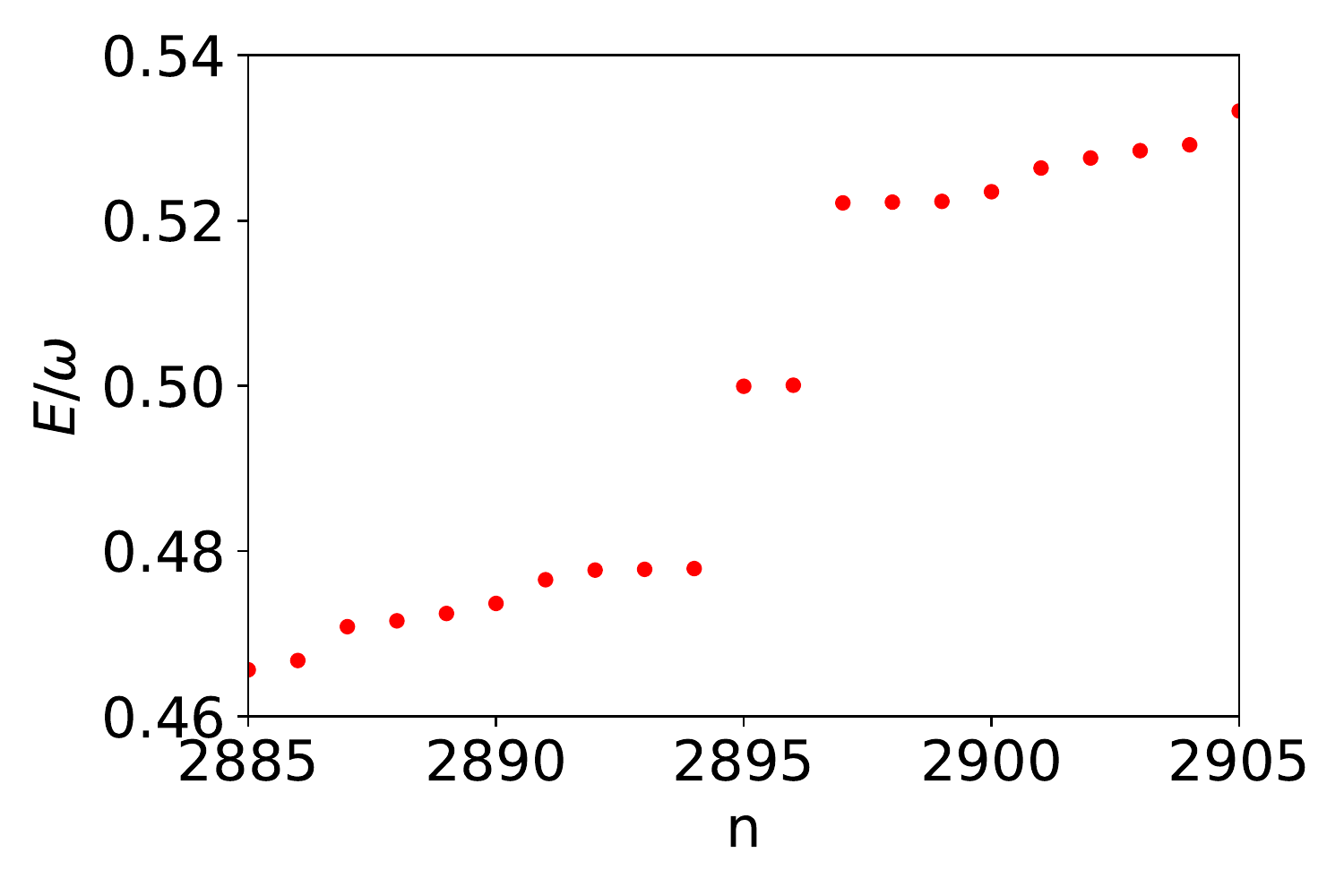}
	\includegraphics[scale=0.29]{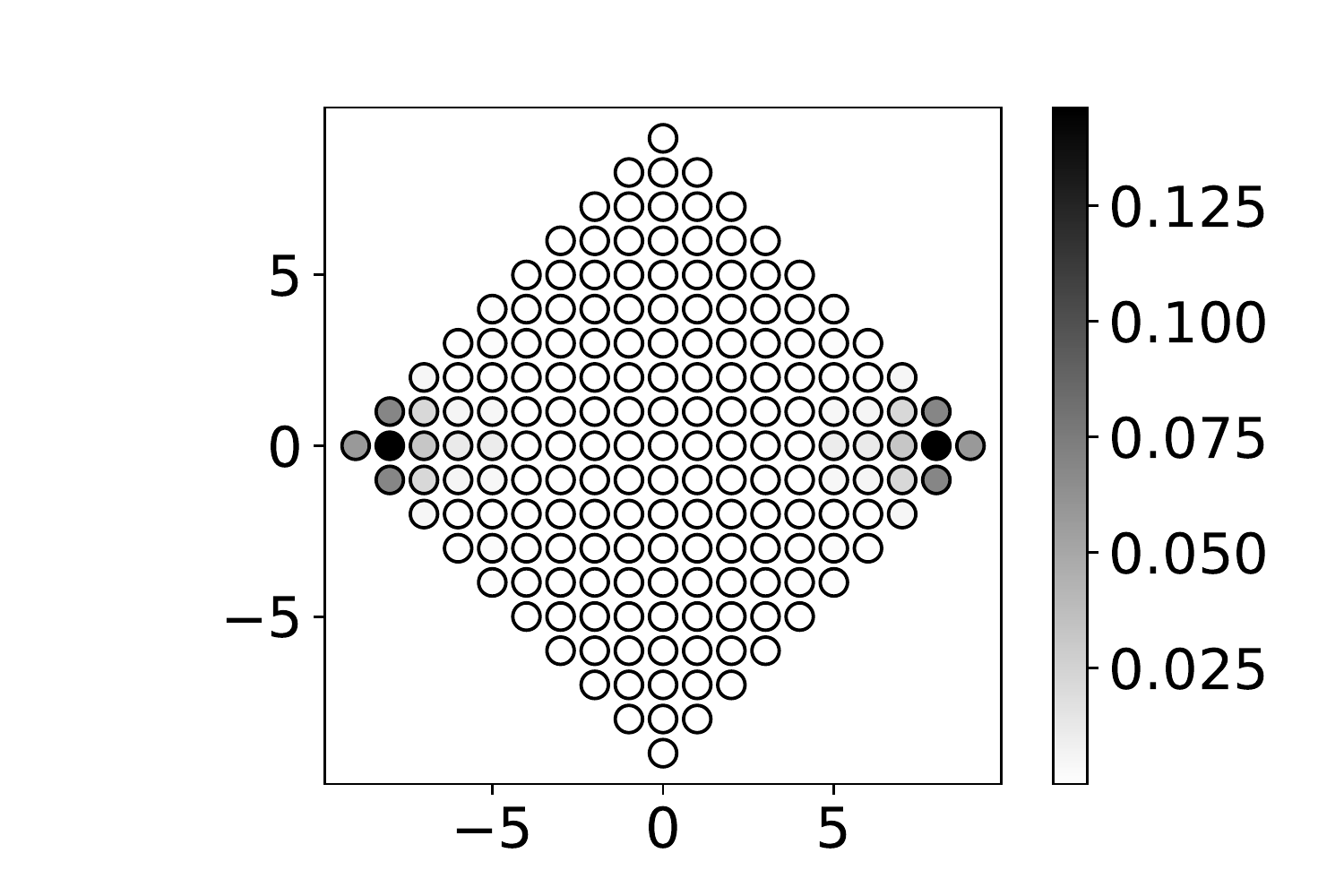}

	\caption{Left Panel: Energy spectrum of the Floquet hamiltonian with $H_1$ of Eq.~\ref{classDH} around quasienergy $\omega/2$ for open boundary conditions in both directions with reflection-symmetry broken edges.  Right panel: Support of the hinge mode for these boundary conditions corresponding to quasienergy $\omega/2$.}
	\label{cornermode}
\end{figure}

This hamiltonian has particle-hole symmetry given by $\mathscr{C}=\sy\ty\rho_x$ and reflection $\mathscr{R}=\sx\rho_z$  where we have introduced a set of Pauli matrices $\rho_{x,y,z}$ for the new spinor degree of freedom in the extended Floquet basis. It belongs to class D and has appropriate commutation relations. Its topological behavior can thus be understood in terms of mirror topological invariant $\mathbb{Z}_2$ (see Supplemental Material). 
The resulting band structure is shown in Fig. \ref{fig:Band-structure-forfinal} with periodic boundary conditions in one direction.  It features gapless modes around quasienergy $\omega/2$ for periodic BC in $y$ direction whenever $\mathbb{Z}_2$ is non-trivial.
Now, if we modify the boundary such that it gives rise to an effective reflection symmetric-breaking mass term, the edge modes are replaced by the hinge modes on reflection-symmetric corners as shown in Fig.~\ref{cornermode}.

{\em A toy model for phonon induced  Rashba SOC-- }
The main ingredients needed for a Rashba SOC are the on-site $\pi$-$\sigma$ spin dependent interactions and the nearest-neighbor $\pi$-$\sigma$ hopping between  same parity orbitals. The spin-dependent on-site  $\pi$-$\sigma$ interaction occurs naturally because of an $\textbf{L}\cdot\textbf{S}$ term, and the nearest neighbor hopping between $\pi$-$\sigma$ orbitals of same parity can be facilitated by the opposite parity orbitals or by ligands. For example, in graphene the intrinsic Rashba SOC occurs because of nearest-neighbor hopping between  $p$ and $s/d$ orbitals~\cite{PhysRevB.82.245412,PhysRevB.74.165310}. 
Furthermore, the ligands can mediate a nearest-neighbor  $\pi$-$\sigma$ hopping which depends on their position, and thus can give rise to a time-dependent Rashba SOC when they oscillate.
\begin{figure}
	\centering
	\includegraphics[scale=0.4]{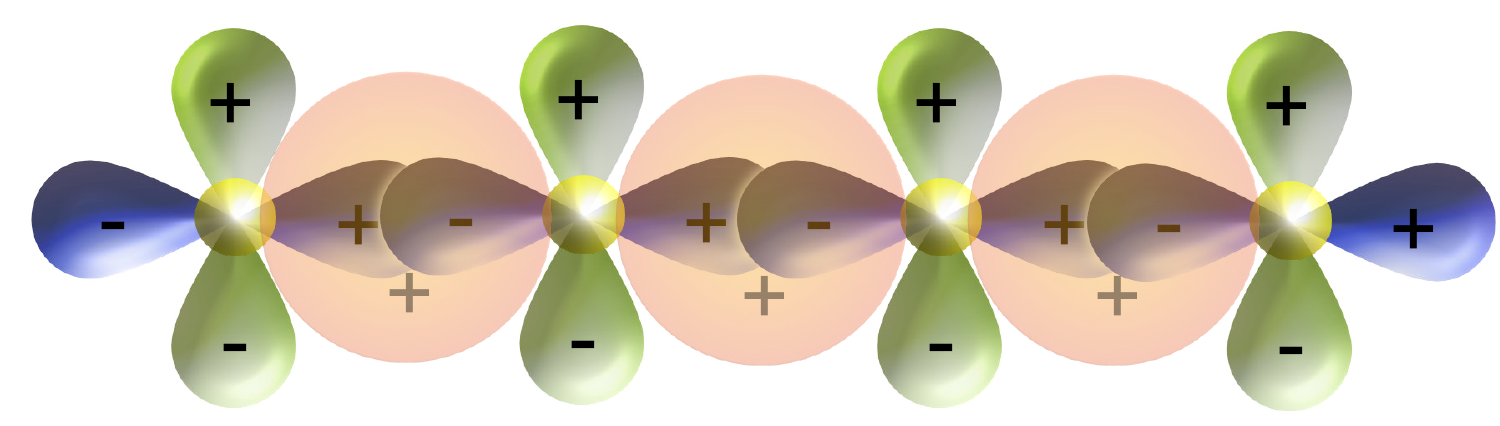}
	\caption{This figure shows the $p_x$ (green) and $p_z$ (blue) orbitals of the two dimensional square lattice along the $x$ direction with $s$ orbitals of the ligands shown in orange color. When the ligand ion is displaced in $z$ direction, it induces a hopping between $p_x$ and $p_z$ orbitals.}
	\label{SOCphonon}
\end{figure}
Consider that each site of our square lattice model considered above has three non-degenerate $p$ orbitals  where neighboring $p_z$ orbitals hybridize to form $\pi$ bands and rest of the orbitals form $\sigma$ bands. These $\pi$ and $\sigma$ orbitals interact via $\textbf{L}\cdot\textbf{S}$ coupling (Table~\ref{LS_table_p}). Furthermore, assume that an additional atom (which we call ligand) with active $s$ orbitals is located between two lattice sites as shown in Fig.~\ref{SOCphonon}.  This arrangement essentially forms a Lieb lattice. When the ligand is displaced in $z$ direction, it induces a hopping between $\pi$ and $\sigma$ orbitals.  When combined with on-site interaction $\textbf{L}\cdot\textbf{S}$ between $\pi$ and $\sigma$ orbitals, it produces an effective spin-dependent hopping between $\pi$ orbitals given by:
\begin{equation}
\begin{split}
\left<p_{z,i,j}|H_{\text{eff}}|p_{z,i+1,j}\right>&=t_{zz}+\frac{1}{\epsilon_{zx}}t_{i,i+1}^{z,x}\left<p_{x,i+1}|\textbf{L}\cdot\textbf{S}|p_{z,i+1}\right>\\&=t_{zz}+i\frac{t_u\zeta}{\epsilon_{zx}}\sigma_y,
\end{split}
\end{equation}
where $t_{zz}=t_0$ is the direct hopping between two $p_z$ orbitals, $\epsilon_{zx}$ is the energy gap between $p_z$ and $p_x$ orbitals, and $t_u$ is the hopping between $p_{z,i}$ and $p_{x,i+1}$ orbital induced by $s$ orbitals of the ligand. 
\begin{table}[h]
	\begin{tabular}{c c c c}
		\hline 
		Orbital & $p_x$ & $p_y$& $p_z$\tabularnewline
		\hline\hline
		$p_x$ & $0$ & $-i\zeta s_z$ & $i \zeta s_y$\tabularnewline
		$p_y$ & $i\zeta s_z$ & $0$ & $-i \zeta s_x$\tabularnewline
		$p_z$ & $-i\zeta s_y$ & $i\zeta s_x$ & $0$\tabularnewline
		\hline
	\end{tabular}
	\caption{Matrix element of $\textbf{L}\cdot\textbf{S}$ operator for  $p$ orbitals.}
	\label{LS_table_p}
\end{table}
Similarly, we can get a $\sigma_x$ dependent hopping in the $y$ direction.  For a small lattice displacement, this ligand-induced hopping
$t_u\approx\left(\frac{u(t)}{L}t_{sp\sigma}\right)t_{sp\sigma}/\epsilon_{sp}$
where $t_{sp\sigma}$ is the hopping between $p_{x/y}$ orbital and the $s$ orbital of the ligand, $\epsilon_{sp}$ is the energy separation between $s$ and $p$ orbitals, $L$ is the distance of the ligand from the neighboring lattice site, and  $t_{sp\sigma}{u(t)}/{L}$ is the hopping between the $\pi$ orbital and  $s$ orbital of the ligand which comes into picture only when the lattice displacement $u(t)$ is non-zero~\cite{PhysRev.94.1498}. 
It gives rise to a time-dependent Rashba SOC which can be controlled by the lattice vibrations associated with the ligand motion. For a coherent phonon, $u(t)\approx u_0\cos\omega t$ where the ratio $u_0/L$ can be as large as $0.1$ in certain cases~\cite{phononamp1,phononamp2}. The ligand-induced hopping $t_u$ depends on a lot of factors like the phonon amplitude,  energy separation $\epsilon_{sp}$, and the hopping $t_{sp\sigma}$ which can be much larger than the $\pi$ hopping $t_{zz}$. Depending on the ligand species, $\epsilon_{sp}\approx1$-$10eV$, and thus the effective hopping $t_{sp\sigma}^2/\epsilon_{sp}$ can be anywhere between $0.1t_{zz}$ and $5t_{zz}$ since the ratio  $t_{sp\sigma}/t_{zz}\approx1$-$10$ usually. This rough estimate indicates that the ligand-induced hopping $t_u$ can be anywhere between $0.1t_{zz}$ and $t_{zz}$, and thus the drive strength $\alpha_0\approx(0.1-1)t_{zz}\zeta/\epsilon_{zx}$ can be of the same order as $t_{zz}$ if the spin-orbit coupling $\zeta$ is comparable to the energy separation between $\pi$ and $\sigma$ orbitals.

{\em 2D system in class AIII.--}
It is known that insulators in class AIII respecting the chiral symmetry (described by $S$) alone have only the trivial band topology. 
When a unitary reflection symmetry (described by $M$, with $M^2=1$) exists, a $\mathbb{Z}$ topological classification 
is possible only when $[S,M]=0$ \cite{Chiu2013, Shiozaki2014}. 

\begin{figure}
\includegraphics[scale=0.23]{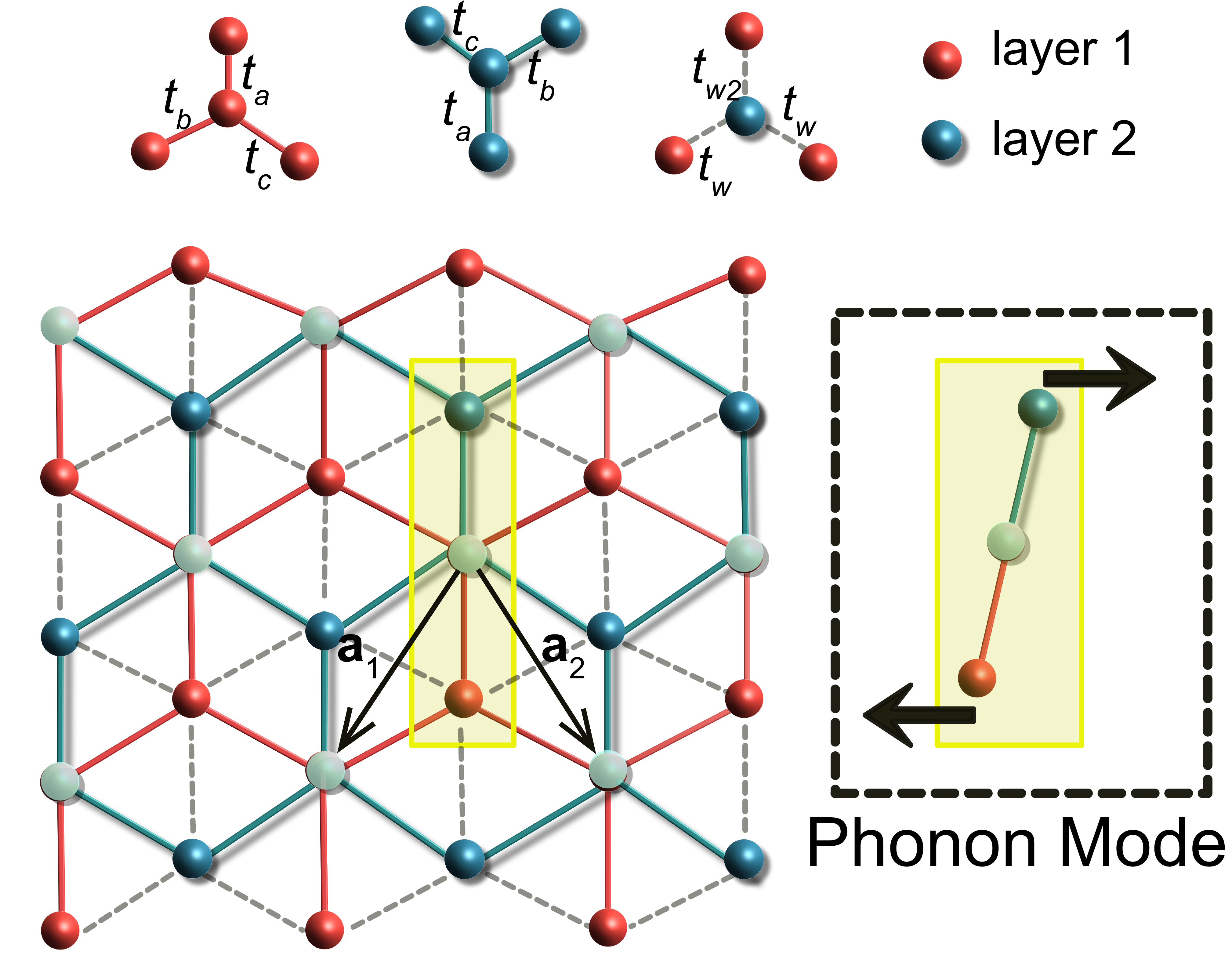}
\caption{A schematic for A-B stacked bilayer honeycomb lattice with nearest-neighbor intra-layer (solid lines) and inter-layer(dashed lines) hopping. The unit cell of the triangular lattice is shown in a yellow  box. The dimer sites (shown in light blue color) don't participate in inter-layer hopping. The right inset shows the phonon mode which affects the hopping parameters $t_b$,$t_c$, and $t_w$.  }
\label{classAIII_model}
\end{figure}

 Consider a tight-binding model for Bernal-stacked bilayer graphene-like lattice with nearest-neighbor intra-layer hopping for all sites and nearest-neighbor inter-layer hopping between non-dimer sites as shown in Fig.~\ref{classAIII_model}.
For periodic boundary conditions, the corresponding Bloch hamiltonian reads
\begin{equation}
\begin{split}
  H_0(\textbf{k})&=t_a\tau_x+2t_b\left(\csxx\csyy\tau_x+\csxx\snyy\ty\right)\\
  &+\left(t_w\csxx\csyy+\frac{t_{w_2}}{2}\csyyt\right)(\sx\tx-\sigma_y\ty)\\
  &+\left(t_w\csxx\snyy+\frac{t_{w_2}}{2}\snyyt\right)(\tx\sigma_y+\ty\sigma_x)\\
  & +t_3(\tx\sx+\sigma_y\tau_y)
\end{split}
\label{h0_half}
\end{equation}
where  $\tau$ and $\sigma$ now operate on sublattice and layer degrees of freedom, respectively. It has chiral symmetry $S=\tz$ and mirror-symmetry $M=\sx\tx$ flipping the coordinates in the $y$ direction.

When a phonon mode is coherently excited
such that the atoms $A_1$ and $B_2$ oscillate out of phase along the 
$x$ direction,
the hopping for the nearest neighbors in $\textbf{a}_1$ and $\textbf{a}_2$ direction as shown in Fig.~\ref{classAIII_model} acquires an additional contribution
\begin{equation}
\begin{split}
H(t)=&\beta(t)\sum_{\textbf{r}_i,\alpha=1,2}\left(a_{\alpha,\textbf{r}_i}^{\dagger}b_{\alpha,\textbf{r}_i+\textbf{a}_1}-a_{\alpha,\textbf{r}_i}^{\dagger}b_{\alpha,\textbf{r}_i+\textbf{a}_2}+h.c\right)\\&+\gamma(t)\sum_{\textbf{r}_i}\left(a_{1,\textbf{r}_i}^{\dagger}b_{2,\textbf{r}_i+\textbf{a}_1}-a_{1,\textbf{r}_i}^{\dagger}b_{2,\textbf{r}_i+\textbf{a}_2}+h.c\right),
\end{split}
\label{drive term1}
\end{equation}
and thus adds
\begin{equation}
\begin{split}
H(\textbf{k},t)=&\beta(t)\left(-\snxx\csyy\ty+\snxx\snyy\tx\right)\\
&+\gamma(t)\snxx\csyy(\sx\ty+\sigma_y\tx)\\&+\gamma(t)\snxx\snyy(\sx\tx-\sigma_y\tau_y)
\end{split}
\end{equation}
to Bloch hamiltonian $H_0(\mathbf{k})$ where $a/b$ refers to $A$ and $B$ sublattice sites, $\alpha$ indicates the layer index, and $\beta(t)$ and $\gamma(t)$ are proportional to the lattice displacement $u(t)$ for small phonon amplitudes. Their magnitude can be estimated as $\beta(t)\approx\eta u(t) t_0/ d_0$,
where $\eta\approx1$-$4$, $t_0$ is the static hopping between two sites, and $d_0$ is the equilibrium separation between
two sites, as the hopping in a tight binding model usually changes 
by a factor of $\left(\frac{d_0}{d_0+u}\right)^\eta$~\cite{PhysRevB.57.6493}. 
For a coherent phonon, the lattice displacement $u(t)\approx u_0\cos\omega t$ which gives $\beta(t)=\beta_0\cos\omega t$, and
thus $ H(\mathbf{k},t)=H_1(\mathbf{k})\cos\omega t$. For a lattice displacement of $5$-$10\%$, the drive strength $\beta_0$ and $\gamma_0$ can be anywhere between $5$-$40\%$ of the static hopping amplitude $t_0$.   

\begin{figure}
	\includegraphics[scale=0.27]{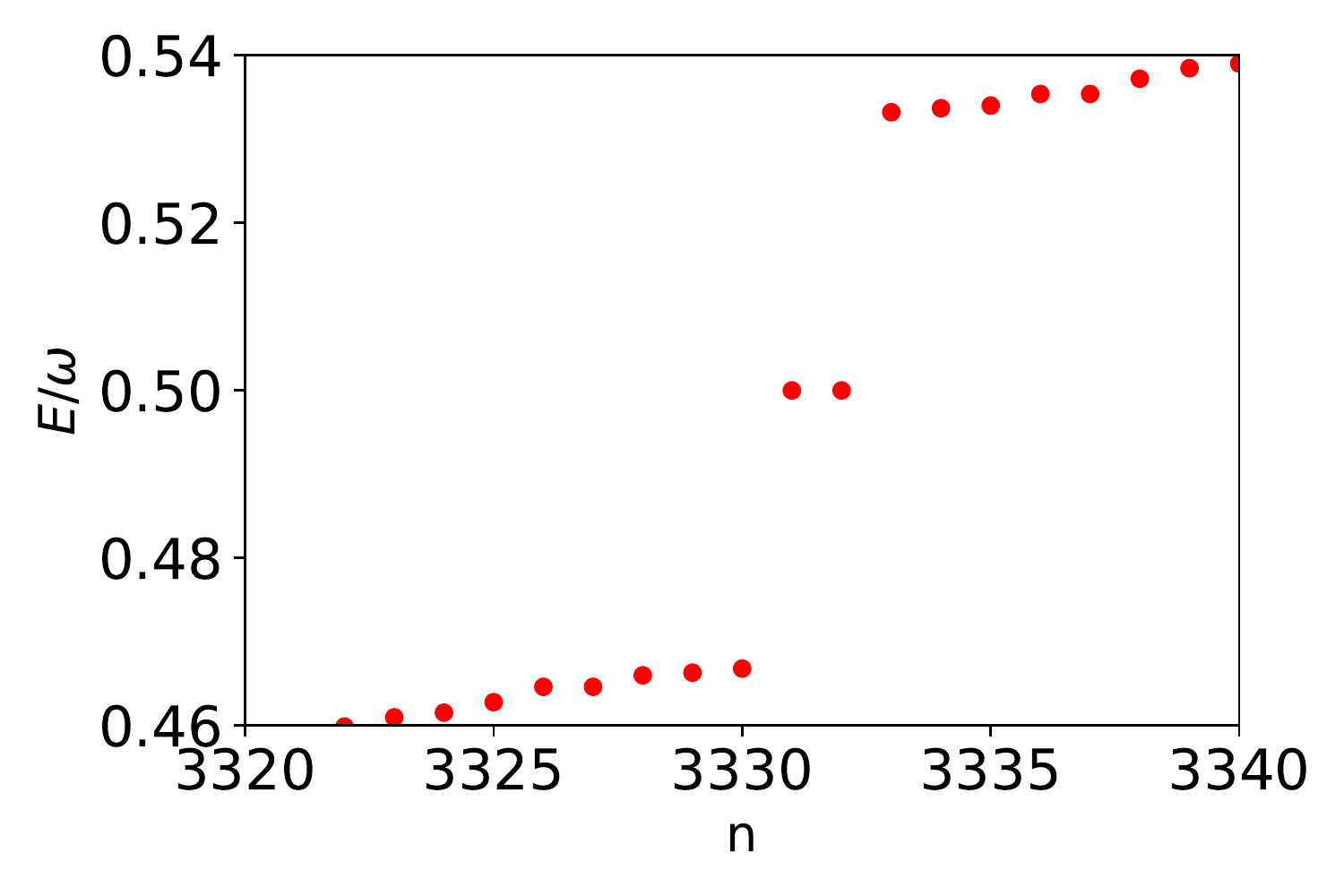}
	\includegraphics[scale=0.27]{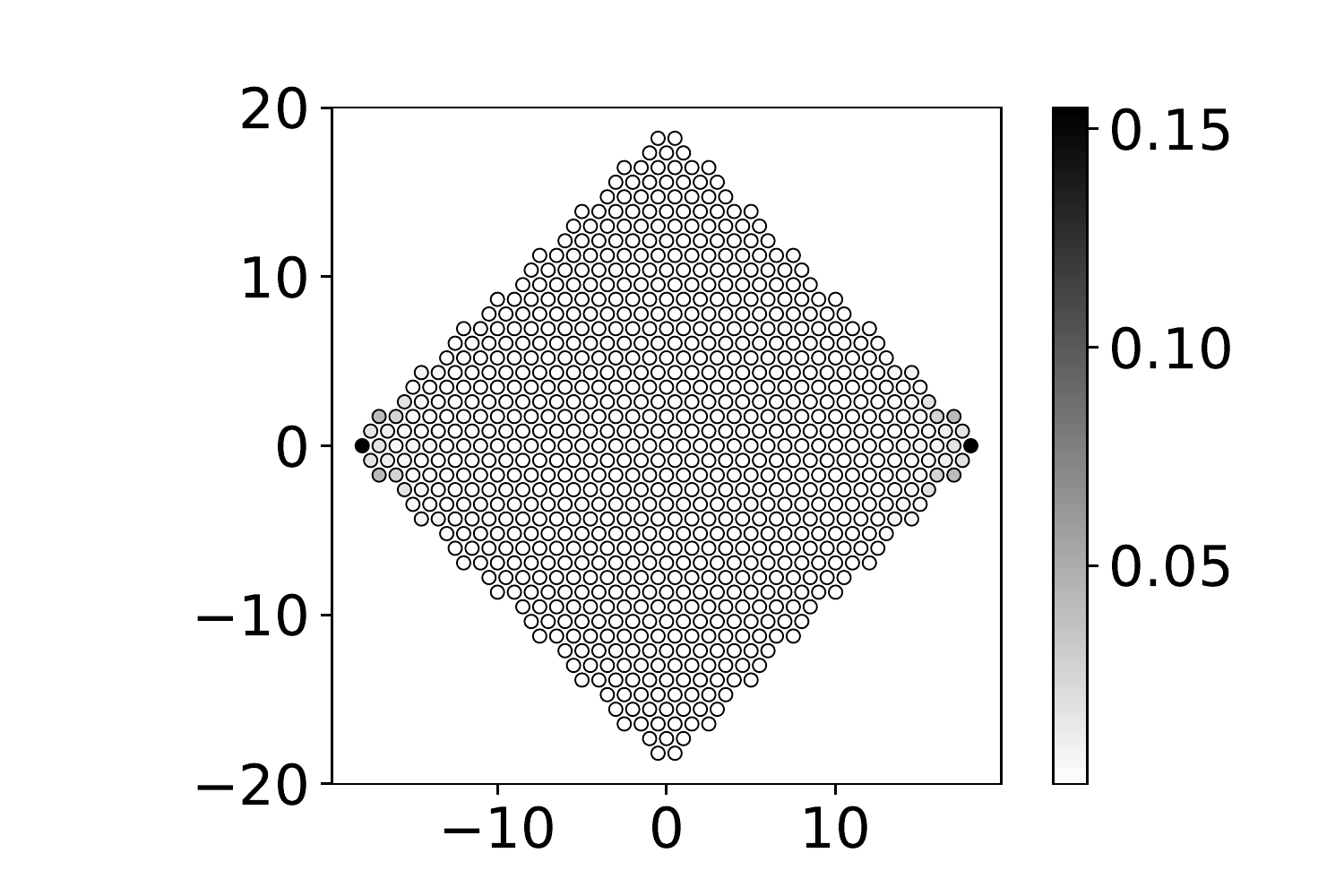}
	\caption{Left panel: Energy spectrum of Floquet hamiltonian with $H_1$ of Eq.\ref{complex} around quasienergy $\omega/2$ for open boundary conditions with reflection-symmetry breaking term. Right panel: Support of the hinge mode for these boundary conditions corresponding to quasinergy $\omega/2$. The parameters for this figure are: $t_a=2.4,t_b=1.2,t_3=0.5,t_w=0.1,\beta_0=0.1,\gamma_0=0.8,\lambda_0=0.8,t_{w_2}=0.1$}
	\label{cornerAIII}

\end{figure}

Now, the two-by-two enlarged hamiltonian (Eq.\ref{eq:Hfl})  has a chiral symmetry, $\mathscr{S}=S\rho_x$ and a reflection symmetry realized by $\mathscr{R}=M\rho_z$. Since $[\mathscr{S},\mathscr{R}]=0$, we can have a nontrivial classification with a mirror $\mathbb{Z}$ topological invariant. In non-trivial regime, it hosts gapless edge modes along $y$ boundaries for the driven system which are protected by reflection-symmetry.
These gapless edge modes at $k_y=0$  co-exist with some some gapless bulk modes at arbitrary $\pm k_y$ which are not protected by the reflection-symmetry. These points arise due to bulk-band gap closings which can be removed, for example, by a drive-induced interlayer imaginary hopping between dimers
\begin{equation}
H_I(t)=\lambda(t)\sum\left(ia_{1,\textbf{r}_i}^{\dagger}b_{2,\textbf{r}_i+\textbf{a}_1}+ia_{1,\textbf{r}_i}^{\dagger}b_{2,\textbf{r}_i+\textbf{a}_2}+h.c\right)
\label{complex}
\end{equation}
which in $k$ space becomes
\begin{equation}
H_I(\mathbf{k})=\lambda(t)\csxx\snyy(\sx\tx+\sigma_y\ty).
\end{equation}
This kind of hopping might not be so easy to realize but it verifies the fact that these gapless bulk modes are not protected by reflection-symmetry alone.
 In this case, if we modify the boundary such that it gives rise to an effective reflection symmetric-breaking mass term, the gapless edge modes are replaced by the hinge modes on reflection-symmetric corners as shown in Fig.\ref{cornerAIII}. Alternatively, the gapless bulk modes can be gapped by translational-symmetry-breaking perturbation, such as a charge density wave. We discuss this possibility in the Supplemental material.

\begin{figure}
	
	\includegraphics[scale=0.27]{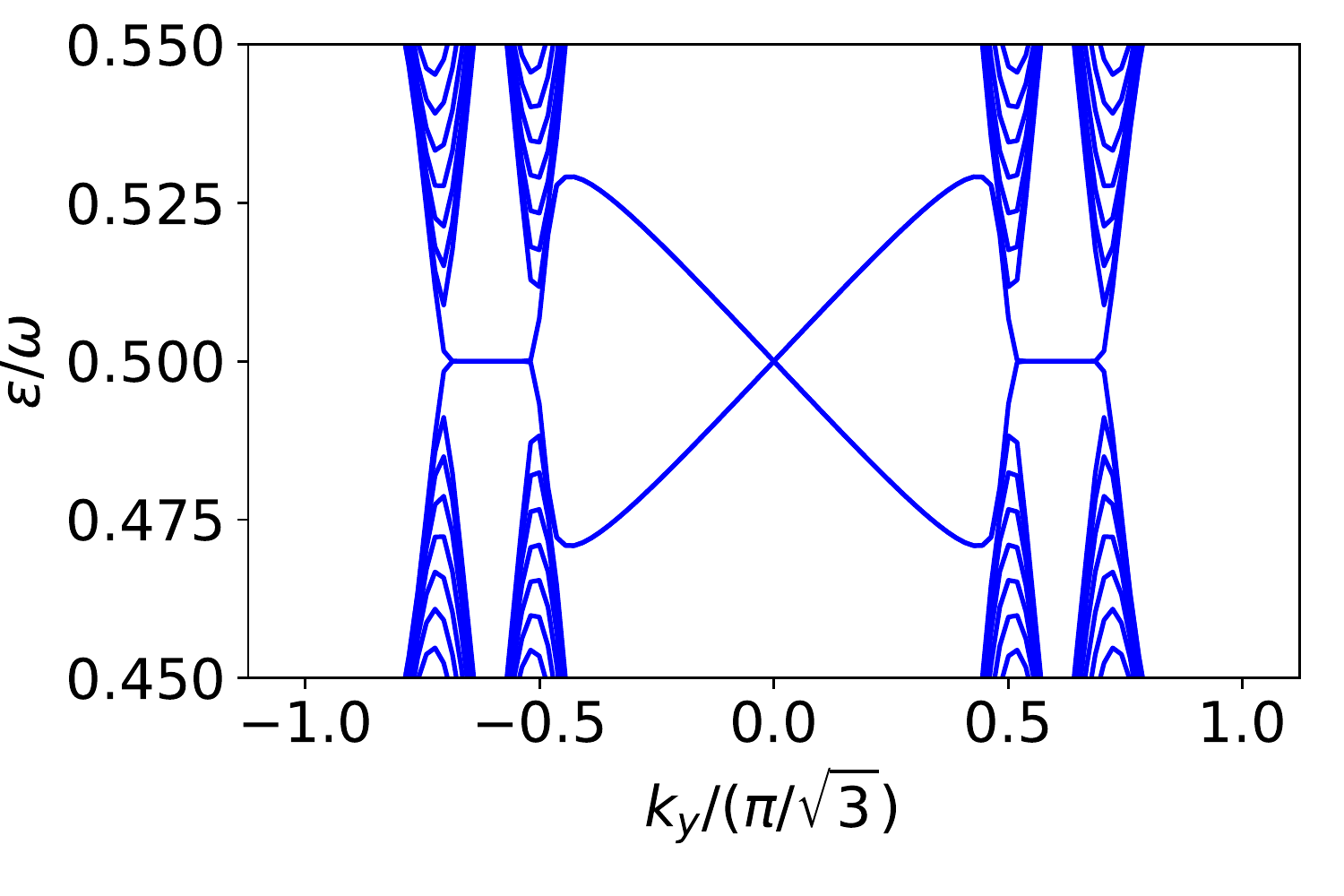}
	\includegraphics[scale=0.27]{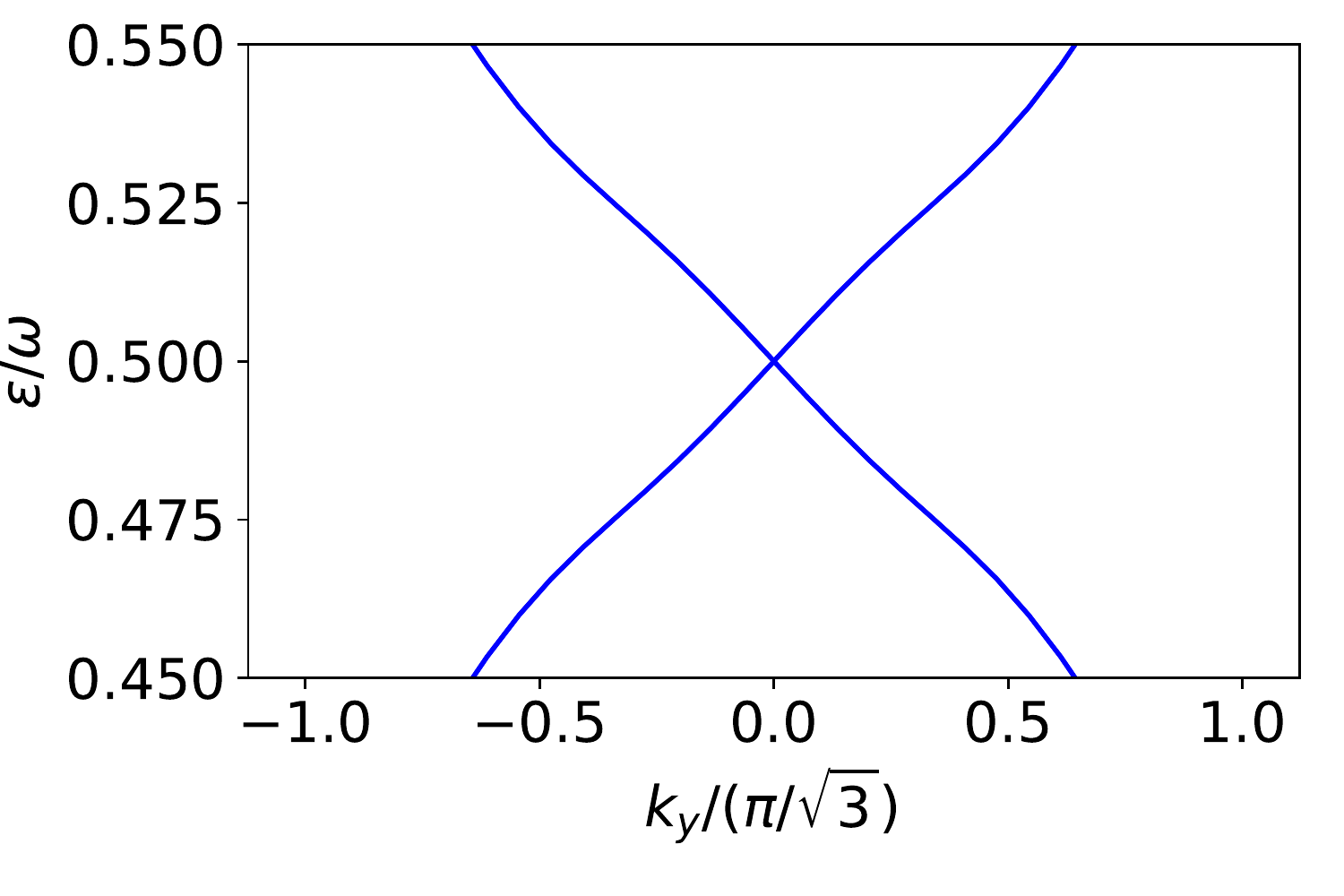}

	\caption{Band structure around $\omega/2$ for  Floquet hamiltonian (upto two Floquet sectors) without/with imaginary hopping (left/right) of Eq.\ref{complex}. At $k_y=0$, time-glide symmetry gives rise to two gapless edge modes which co-exist with other gapless modes in the absence of imaginary hopping term. The parameters for left figure are: $t_a=2.4,t_b=1.2,t_3=0.5,\beta_0=0.1,t_w=0.5,\gamma_0=0.5,\omega=4.4,t_{w_2}=0.1$ and for the right figure $\gamma_0=0.8,\lambda=0.8$.}
	\label{BLG_floquetzoneedge}
\end{figure}

{\em Conclusions.-- }
We discussed how a phonon drive can be used to promote a static symmetry to a space-time symmetry which in turn can allow the system to exhibit a topological classification.  Both of these schemes need phonon frequencies to be of the same order as hopping parameter. These phonon frequencies usually depend on the bond strength while the hopping parameter in graphene-like lattices depend on overlap between neighboring $\pi$ orbitals. This kind of parameter regime can be realized by suppressing the hopping parameter without affecting the bond strengths between neighboring atoms which determine the phonon frequencies. Alternatively, one can consider placing the system on a substrate which binds to different sites such that the phonon frequencies are increased. These lattice models can be possibly realized using covalent organic framework~\cite{cof1,liebcof} where molecular orbitals play the same role as atomic orbitals in our model and hopping parameters are of the order of 100meV. Similarly, twisted bilayer materials can provide another platform to realize the Class AIII model where hopping parameters can be made comparable to certain phonon frequencies~\cite{TLGmain, TLgphonon}.

{\em Acknowledgement.---} We acknowledge support from the Institute of Quantum Information and Matter, 
an NSF Frontier center funded by the  Gordon  and  Betty  Moore  Foundation,  the  Packard
Foundation, and the Simons foundation.
A. H. and Y.P. are grateful to support from the Walter Burke Institute for Theoretical Physics at Caltech.
G. R. is grateful to the support from the  ARO  MURI  W911NF-16-1-0361 Quantum Materials by 
Design with Electromagnetic Excitation”  sponsored  by  the  U.S.  Army. 
\bibliographystyle{apsrev4-1} 
\bibliography{draft}



\newpage

\onecolumngrid
\section*{Supplementary material for ``Phonon-induced Floquet second-order topological phases protected by space-time symmetries''}
\section{Effective Symmetry Operators and modified commutation relations in
	extended Floquet basis}

In this section, we briefly review the effective symmetry operations
for the frequency-domain representation of the time-periodic hamiltonian
when the system has a time-glide symmetry.

In this case, we show that the original time-glide symmetry of a time-periodic
hamiltonian is mapped to the reflection symmetry for the effective
hamiltonian~\cite{Peng2019}. The time-periodic
hamiltonian here follows:
\begin{equation}
MH(\mathbf{k},t)M^{-1}=H(-k_{x},\mathbf{k}_{\parallel},t+T/2).
\end{equation}Here, we are interested in studying these systems in fourier basis
and thus consider the above hamiltonian in extended Floquet basis
given by:
\begin{equation}
H_{\text{full}}(\mathbf{k})=\begin{pmatrix}.\\
& H_{0}+2\omega & H_{1} & H_{2} & . & .\\
& H_{\bar{1}} & H_{0}+\omega & H_{1} & H_{2}\\
& H_{\bar{2}} & H_{\bar{1}} & H_{0} & H_{1} & H_{2} & .\\
& . & H_{\bar{2}} & H_{\bar{1}} & H_{0}-\omega & H_{1} & H_{2} & .\\
&  &  & H_{\bar{2}} & H_{\bar{1}} & H_{0}-2\omega & H_{1} & H_{2} & .\\
&  &  &  &  &  & .\\
&  &  &  &  &  &  & .\\
\\
\end{pmatrix},
\end{equation}
and 
\begin{equation}
H_{n}(\mathbf{k})=\frac{1}{T}\int_{0}^{T}H(\mathbf{k},t)e^{-in\omega t}.
\end{equation}
In this basis, the time-glide symmetry is given by effective reflection
\begin{equation}
\mathcal{R}_{\text{eff}}=\begin{pmatrix}.\\
& M\\
&  & -M\\
&  &  & M\\
&  &  &  & -M\\
&  &  &  &  & .\\
\\
\end{pmatrix},
\end{equation}and other symmetry operator are also modified, for example the new
(effective) operators for time-reversal, charge-conjugation, and chiral
symmetry are now given by:
\begin{equation}
\mathscr{T}=\begin{pmatrix}\\
& \mathcal{T}\\
&  & \mathcal{T}\\
&  &  & \mathcal{T}\\
&  &  &  & \mathcal{T}\\
&  &  &  &  & \mathcal{T}\\
\\
\end{pmatrix},\ \mathscr{C}=\begin{pmatrix}\\
& \mathcal{}\\
&  & \mathcal{} &  & \mathcal{C}\\
&  &  & \mathcal{C}\\
&  & \mathcal{C} &  & \mathcal{}\\
&  &  &  &  & \mathcal{}\\
\\
\end{pmatrix},\mathscr{S}=\begin{pmatrix}\\
& \mathcal{}\\
&  & \mathcal{} &  & S\\
&  &  & \mathcal{S}\\
&  & \mathcal{S} &  & \mathcal{}\\
&  &  &  &  & \mathcal{}\\
\\
\end{pmatrix}.
\end{equation}Now, the commutation relation between effective different symmetry
operators is different from the commutation relation for the static
case. This indicates that the effective hamiltonian although belongs
to the same AZ class as the static hamiltonian but can allow the existence
of a non trivial topological phase (For e.g $\{M,S\}=0\Longrightarrow[\mathcal{R},\mathscr{S}]=0$)
in the presence of this time-glide symmetry. We are going to exploit
this feature in realizing non-topological phase by altering the
modified commutation relations which results in a change in topological classification as shown in Table.~\ref{table2}. Although, in this work we use a periodic drive to promote reflection symmetry to a time-glide symmetry in two dimensions, the same ideas can be applied to two-fold rotations and inversion symmetry in two and three dimensions and the consequences of such promotions are discussed in table ~\ref{table2} and table~\ref{table3}.

\begin{table}[h]
\begin{tabular}{|c|c|c|c|c|}
	\hline 
	\multirow{2}{*}{Class} & \multicolumn{2}{c|}{2D} & \multicolumn{2}{c|}{3D}\tabularnewline
	\cline{2-5} \cline{3-5} \cline{4-5} \cline{5-5} 
	& Symmetry Promotion & Classification & Symmetry Promotion & Classification\tabularnewline
	\hline 
	AIII & $R_{-}$ & $\mathbb{Z}$ & $R_{+}$ & $\mathbb{Z}^{2}$\tabularnewline
	\hline 
	BDI & $R_{+-}$ & $\mathbb{Z}$ & $R_{--}$ & $\mathbb{Z}$\tabularnewline
	\hline 
	D & $R_{-}$ & $\mathbb{Z}_{2}$ & $R_{-}$ & $\mathbb{Z}$\tabularnewline
	\hline 
	DIII & $R_{+-}$ & $\mathbb{Z}_{2}$ & $R_{--}$ & $\mathbb{Z}^{2}$\tabularnewline
	\hline 
	CII & $R_{+-}$ & $2\mathbb{Z}$ & $R_{++}$ & $\mathbb{Z}_{2}^{2}$\tabularnewline
	\hline 
	CI & $R_{-+}$ & $2\mathbb{Z}$ & $R_{++}$ & $\mathbb{Z}$\tabularnewline
	\hline 
	\hline 
	AIII & $C_{2+}$ & $\mathbb{Z}$ & $C_{2-}$ & $\mathbb{Z}^{2}$\tabularnewline
	\hline 
	BDI & $C_{2++}$ & $\mathbb{Z}$ & $C_{2+-}$ & $\mathbb{Z}$\tabularnewline
	\hline 
	DIII & $C_{2++}$ & $\mathbb{Z}$ & $C_{2+-}$ & $\mathbb{Z}$\tabularnewline
	\hline 
	CII & $C_{2++}$ & $2\mathbb{Z}$ & $C_{2+-}$ & $\mathbb{Z}$\tabularnewline
	\hline 
	CI & $C_{2++}$ & $2\mathbb{Z}$ & $C_{2+-}$, $C_{2-+}$ & $\mathbb{Z}$, $(2\mathbb{Z})^{2}$\tabularnewline
	\hline 
\end{tabular}
\caption{Symmetry promotion from reflection $R$ and two-fold rotation $C$.}
\label{table2}
\end{table}
\begin{table}[h!]
\begin{tabular}{|c|c|c|}
	\hline 
	\multirow{2}{*}{Class} & \multicolumn{2}{c|}{3D}\tabularnewline
	\cline{2-3} \cline{3-3} 
	& Symmetry Promotion & Classification\tabularnewline
	\hline 
	AIII & $I_{+}$ & $\mathbb{Z}^{2}$\tabularnewline
	\hline 
	BDI & $I_{++}$ & $\mathbb{Z}$\tabularnewline
	\hline 
	D & $I_{+}$ & $\mathbb{Z}$\tabularnewline
	\hline 
	DIII & $I_{++}$ & $\mathbb{Z}^{2}$\tabularnewline
	\hline 
	CII & $I_{--}$ & $\mathbb{Z}_{2}^{2}$\tabularnewline
	\hline 
	CI & $I_{--}$, $I_{++}$ & $\mathbb{Z}$, $(2\mathbb{Z})^{2}$\tabularnewline
	\hline 
\end{tabular}
\caption{Symmetry promotion from inversion $I$}
\label{table3}
\end{table}

When the drive is monochromatic, $H_n=0$ for $|n|>1$, and thus the above symmetries can be best described in terms of constraints on $H_0$ and $H_1$. In this case, the chiral symmetric hamiltonians satisfy:
\begin{equation}
SH_{0}(\mathbf{k},\mathbf{r})S^{-1}=-H_{0}(\mathbf{k},\mathbf{r}),\quad SH_{1}(\mathbf{k},\mathbf{r})S^{-1}=-H_{1}^{+}(\mathbf{k},\mathbf{r}).
\end{equation}
Similarly particle-hole symmetry is given by:
\begin{equation}
CH_{0}^{*}(\mathbf{k},\mathbf{r})C^{-1}=-H_{0}(-\mathbf{k},\mathbf{r}),\quad CH_{1}^{*}(\mathbf{k},\mathbf{r})C^{-1}=-H_{1}^{+}(-\mathbf{k},\mathbf{r})
\end{equation}
and time-glide by:
\begin{equation}
MH_{0}(k_{x},k_{y})M^{-1}=H_{0}(-k_{x,}k_{y}),\quad MH_{1}(k_{x},k_{y})M^{-1}=-H_{1}(-k_{x,}k_{y}).
\end{equation}

\section{Mirror topological invariant $M\mathbb{Z}_2$ for class D, $R_{+}$ in two dimensions}
\begin{figure}[h]
	\includegraphics[scale=0.2]{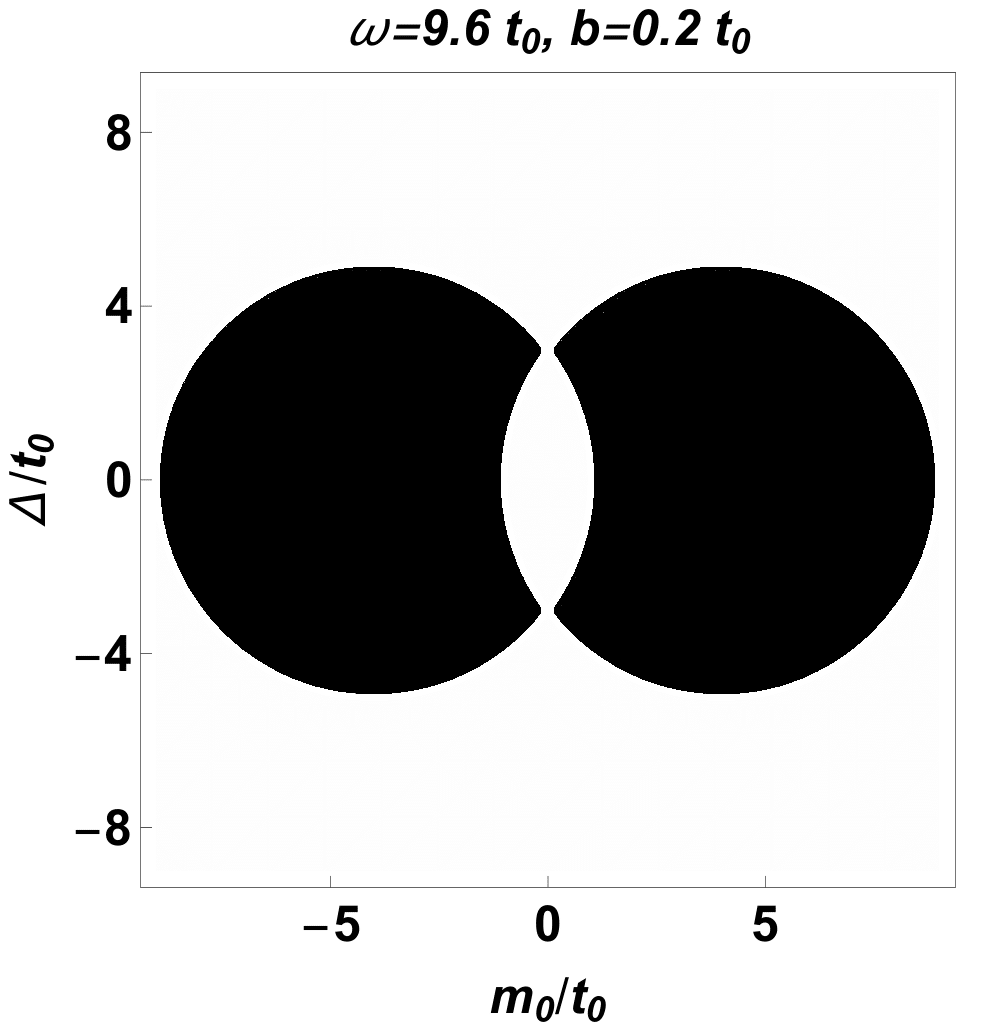}
		\includegraphics[scale=0.2]{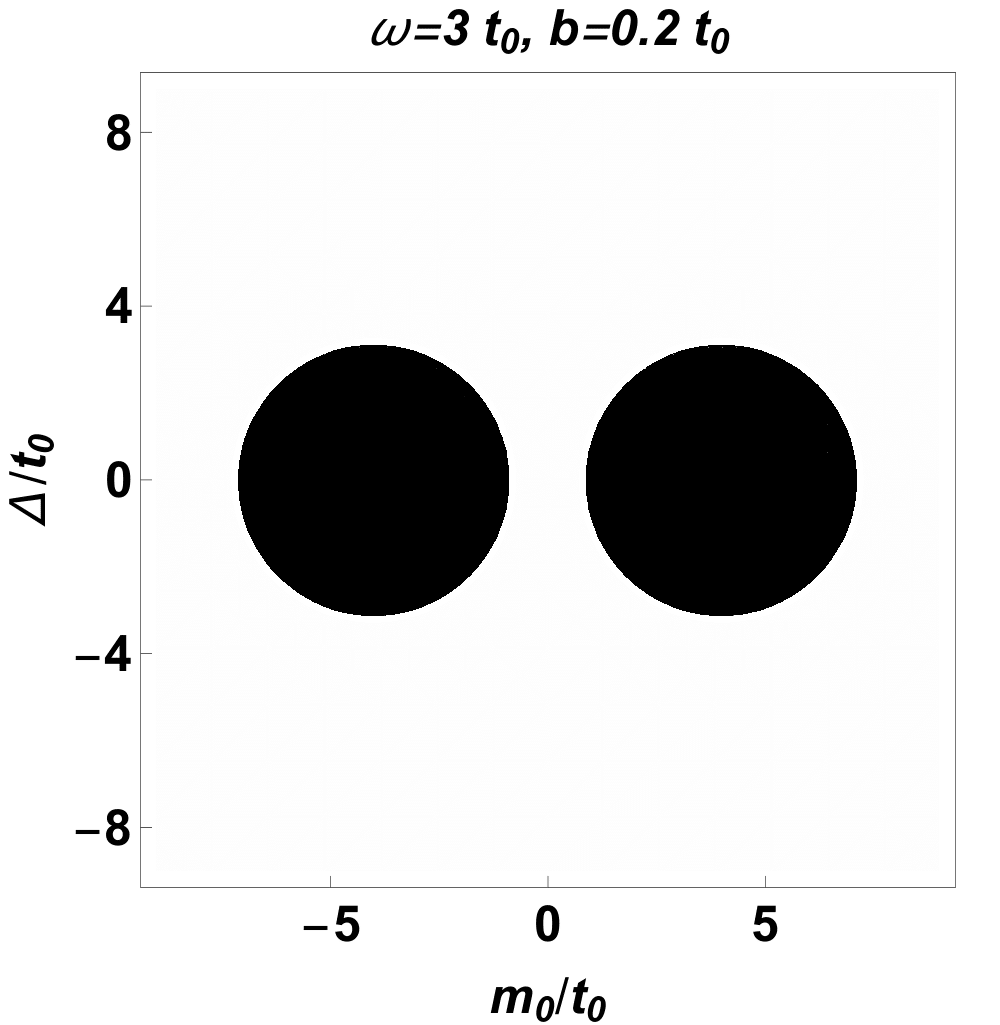}
	\caption{Mirror topological invariant $\eta_{M\mathbb{Z}_{2}}$ on parameters
		$m_{0}$ and $\Delta$ for two different values of $\omega$. All parameters are given in units of hopping amplitude $t_0$. Black regions
		indicate the non-trivial region.\label{fig:Mirror-topological-invariant}}
\end{figure}

\begin{figure}[h]
	\includegraphics[scale=0.45]{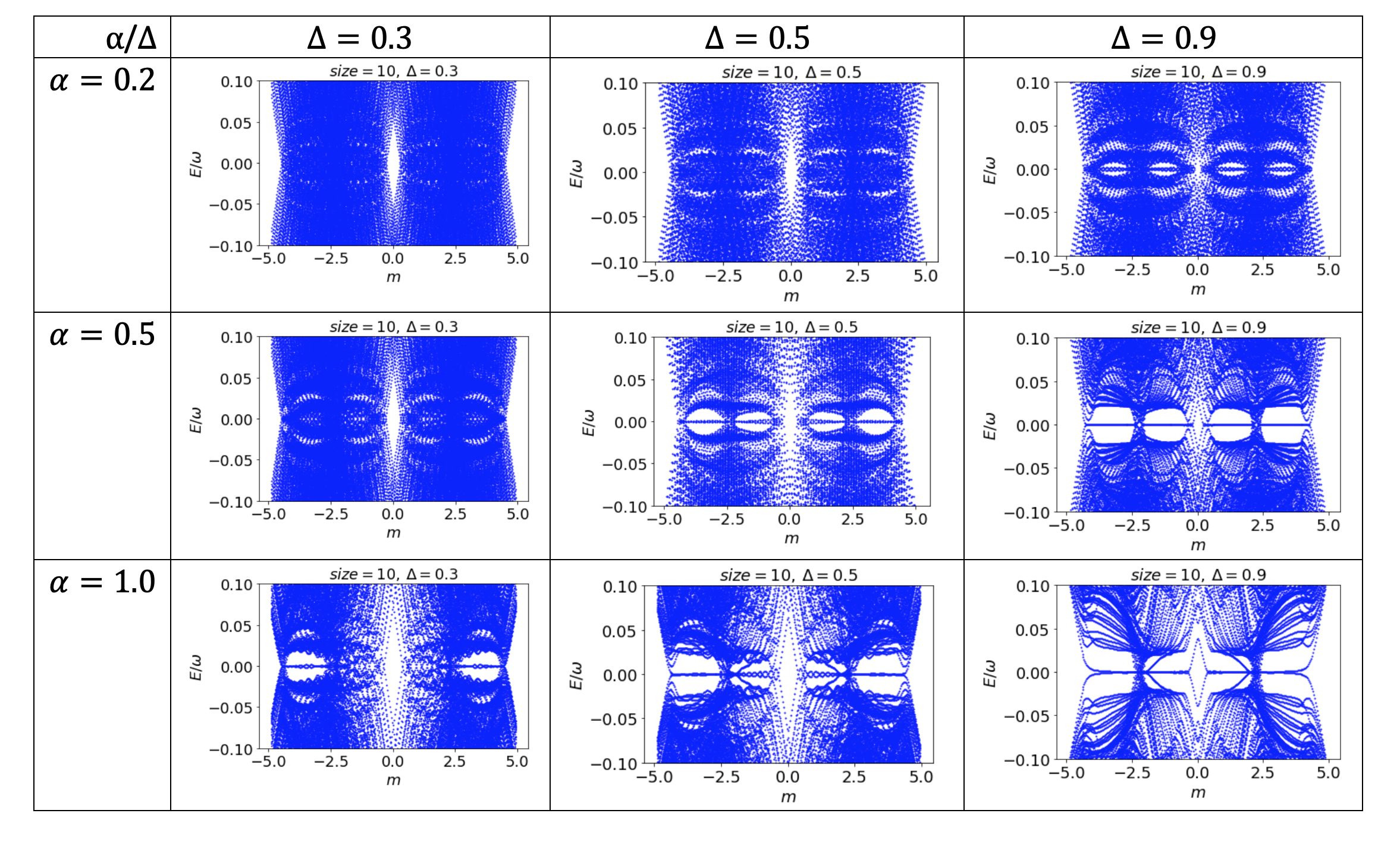}\caption{Energy spectrum around zero energy as a function of $m$ for an eight band model capturing the main features of the $\tau_z$ drive considered in Eq.~\ref{eq:fullfloquet1}.}
	\label{eightbandtz}
\end{figure}

For class D, when the particle-hole and reflection symmetry operator commute, the different topological phases  can be distinguished on the basis of mirror topological invariant $M\mathbb{Z}_2$. This invariant can be calculated at reflection symmetric hyperplanes by first
block diagonalizing the hamiltonian in $\mathscr{R}$ basis, and then
calculating the $\mathbb{Z}_2$ invariant for one block. When the reflection operator commutes with the effective particle-hole operator $\mathscr{C}$,  these two blocks do not mix and hence a classification can be made on the basis of this topological invariant for one block. In the main text we considered the undriven model:
\begin{equation}
H_{0}(k_{x},k_{y})=m_1\tau_{z}+\Delta\tau_{x}+b\sigma_{x}
\end{equation}
where $m_1=m-2t_0\cos k_{x}-2t_0\cos k_{y}$ with a drive of the form:
\begin{equation}
H(t)=2\alpha_0\cos\omega t(\snx\sigma_y-\sny\sigma_x)\tau_z=H_1e^{i\omega t}+H_1^\dagger e^{-i\omega t}.
\end{equation}
This hamiltonian has a particle-hole symmetry given by $C=\ty\sy$ and time-glide with reflection $M=\sx$ about y axis ($y\rightarrow-y$).
We can cast it into a more familiar form if we use the eigenstate basis of $m(\textbf{k})\tz+\Delta\tx$ which corresponds to a transformation  $\tz\rightarrow\cos\theta\tz-\sin\theta\tx$
where $\cos\theta=\frac{m(\textbf{k})}{\sqrt{m(\textbf{k})+\Delta^{2}}}.$ In this basis:
\begin{equation}
H_{0} = \sqrt{m(\textbf{k})^{2}+\Delta^{2}}\tz+b\sx,
\end{equation} 
and
\begin{equation} H_1=2\alpha_{0}(\snx\sy-\sny\sx)\left(\frac{m(\textbf{k})}{E_{k}}\tz-\frac{\Delta}{E_{k}}\tx\right)
\end{equation}
where $E_k=\sqrt{m(\textbf{k})^2+\Delta^2}$. In this basis, the particle-hole operator $C=\ty\sy$ and time-glide  $M=\sx$ remains the same. Now the Floquet hamiltonian for photon sectors $n$ and $n+1$ can be written as:
\begin{equation}
H_F(\mathbf{k})=\begin{pmatrix}E_{k}\tz+\frac{\omega}{2}\mathbb{I}+b\sx & \alpha_{0}(\snx\sy-\sny\sx)\left(\frac{m(\textbf{k})}{E_{k}}\tz-\frac{\Delta}{E_{k}}\tx\right)\\
\alpha_{0}(\snx\sy-\sny\sx)\left(\frac{m(\textbf{k})}{E_{k}}\tz-\frac{\Delta}{E_{k}}\tx\right) & E_{k}\tz-\frac{\omega}{2}\mathbb{I}+b\sx
\end{pmatrix}.
\label{eq:floquetD}
\end{equation}
The topological behavior of this hamiltonian can also be understood by focusing on its inner $4\times4$ block given by:
\begin{equation}
\mathscr{H}(\mathbf{k})=\begin{pmatrix}(-E_{k}+\frac{\omega}{2})\mathbb{I}+b\sx & \alpha_{0}(\snx\sy-\sny\sx)\frac{\Delta}{E_{k}}\\
\alpha_{0}(\snx\sy-\sny\sx)\frac{\Delta}{E_{k}} & (E_{k}-\frac{\omega}{2})\mathbb{I}+b\sx
\end{pmatrix}=(-E_{k}+\frac{\omega}{2})\rho_z+b\sigma_x+\alpha_{0}(\snx\sy-\sny\sx)\frac{\Delta}{E_{k}}\rho_x
\label{eq:inner-2}
\end{equation}where $\rho$ indicates the photon degree of freedom. In this case, the particle-hole symmetry $\mathscr{C}=\sy\py$ and it has a reflection symmetry $\mathscr{R}=\sx\rho_z$ which commutes with $\mathscr{C}$, and hence it can
be characterized by a mirror $\mathbb{Z}_{2}$ invariant. In order to calculate this topological invariant we go to reflection
symmetric hyperplanes ($k_{y}=0,\pi$) and express this hamiltonian
in the basis of M arranged such that first block has eigenvalue $+1$.
The above hamiltonian of Eq. \ref{eq:inner-2} takes the following form:
\begin{equation}
\mathscr{H}(k_{x})=\begin{pmatrix}H_{+}\\
& H_{-}
\end{pmatrix}=\begin{pmatrix}-E_{k}+\frac{\omega}{2}+b & -i\alpha_{0}\snx\frac{\Delta}{E_{k}} & 0 & 0\\
i\alpha_{0}\snx\frac{\Delta}{E_{k}} & E_{k}-\frac{\omega}{2}-b & 0 & 0\\
0 & 0 & E_{k}-\frac{\omega}{2}+b & -i\alpha_{0}\snx\frac{\Delta}{E_{k}}\\
0 & 0 & i\alpha_{0}\snx\frac{\Delta}{E_{k}} & -E_{k}+\frac{\omega}{2}-b
\end{pmatrix}
\end{equation}
with
\begin{equation}
R=\begin{pmatrix}1\\
& 1\\
&  & -1\\
&  &  & -1
\end{pmatrix}\,\text{and }P=\begin{pmatrix} & -1\\
-1\\
&  &  & 1\\
&  & 1
\end{pmatrix}
\end{equation} in the new basis. Now, each block belongs to class D and in order to calculate the mirror
$\mathbb{Z}_{2}$ invariant we can pick any block. For example if
we pick $H_{+},$ then the invariant can be calculated from Pffafin
of $H_{+}$ at $k_{x}=0,\pi$. At $k_{y}=0,$ the $\mathbb{Z}_{2}$
topological invariant is given by:
\begin{equation}
\eta_{k_{y}=0}=\text{Sign}[(E_{k}-\frac{\omega}{2}+b)_{(k_{x}=0,k_{y}=0)}(E_{k}-\frac{\omega}{2}+b)_{(k_{x}=\pi,k_{y}=0)}].
\end{equation}Similarly, at $k_{y}=\pi$
\begin{equation}
\eta_{k_{y}=\pi}=\text{Sign}[(E_{k}-\frac{\omega}{2}+b)_{(k_{x}=0,k_{y}=\pi)}(E_{k}-\frac{\omega}{2}+b)_{(k_{x}=\pi,k_{y}=\pi)}],
\end{equation}and thus the mirror $\mathbb{Z}_{2}$ invariant is now given by :
\begin{equation}
\eta_{M\mathbb{Z}_{2}}=1-|\eta_{k_{y}=0}-\eta_{k_{y}=\pi}|=\text{Sign}[\eta_{k_{y}=0}\eta_{k_{y}=\pi}],
\end{equation}and its dependence on $m_{0}$ and $\Delta$ is shown in Fig. \ref{fig:Mirror-topological-invariant}. This indicates that if the system size is quite large, we can get a non-trivial phase for a very small value of superconducting gap $\Delta$ but for small system sizes we find the gapless edge modes only for $\Delta\approx0.5$ as the bulk gap becomes very small for lower values of $\Delta$.
Although, the topological invariant is non-trivial for a large range of $m$ and $\Delta$ but the system does not seem  to exhibit corner modes for small values of $\Delta$. In order to understand the regime for gapless boundary states we study an eightband model which captures the essential features of the above model. This model is given by hamiltonian:
\begin{equation}
H_{\text{eff}}=\begin{pmatrix}H_{0}+\frac{\omega}{2}&{H_1}\\
H_1^{\dagger}& H_{0}-\frac{\omega}{2}.
\end{pmatrix}
\label{eq:fullfloquet1}
\end{equation}
The spectrum of this hamiltoninan is shown in  Fig.~\ref{eightbandtz} for open boundary conditions. In certain cases, the bulk gap becomes very small and thus the gapless boundary modes or zero energy hinge modes can not be observed for a small system size.

\section{Effect of translation-symmetry breaking perturbations on gapless bulk modes at Floquet zone boundaries}
\begin{figure}[h]
	\includegraphics[scale=0.5]{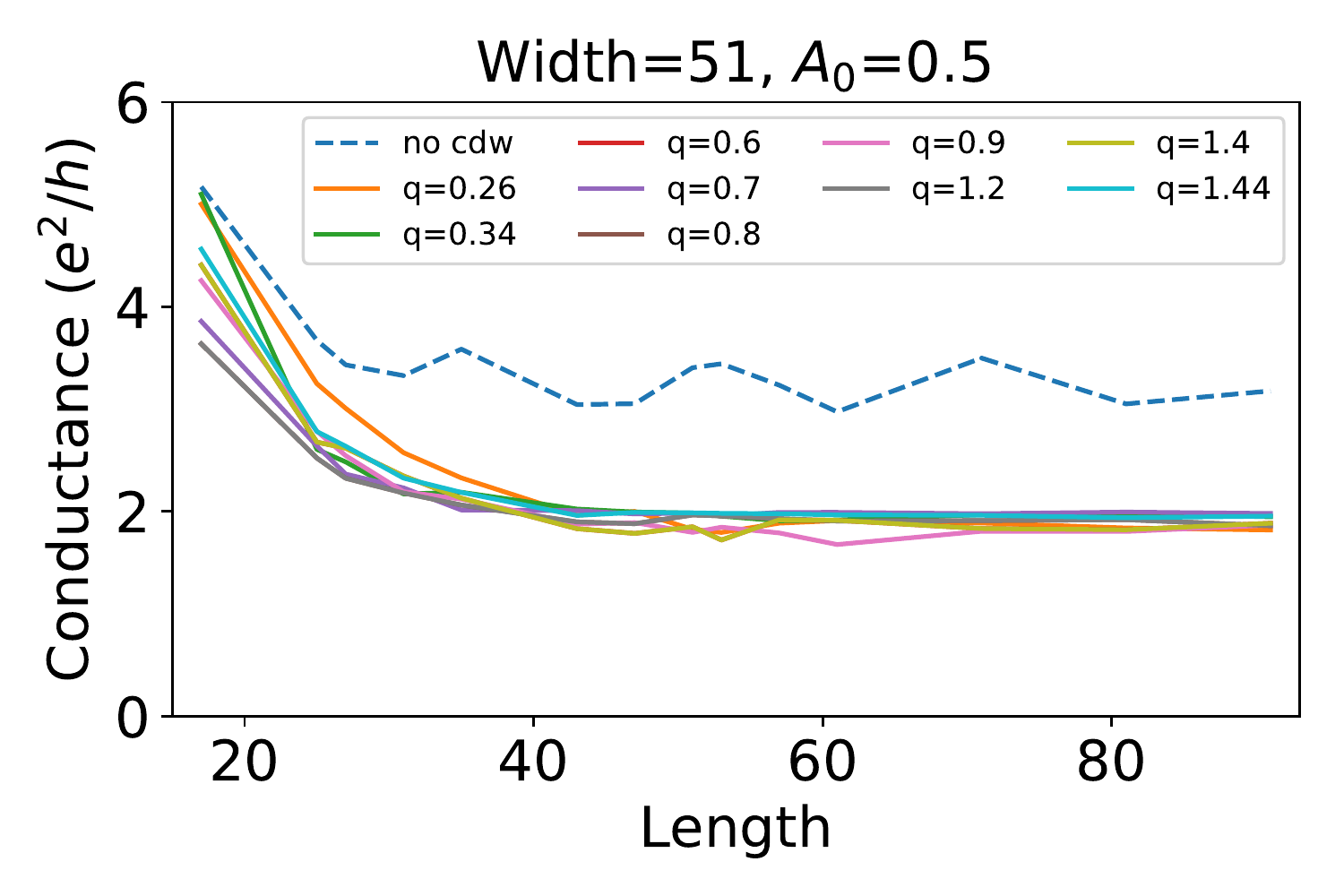}
	\includegraphics[scale=0.5]{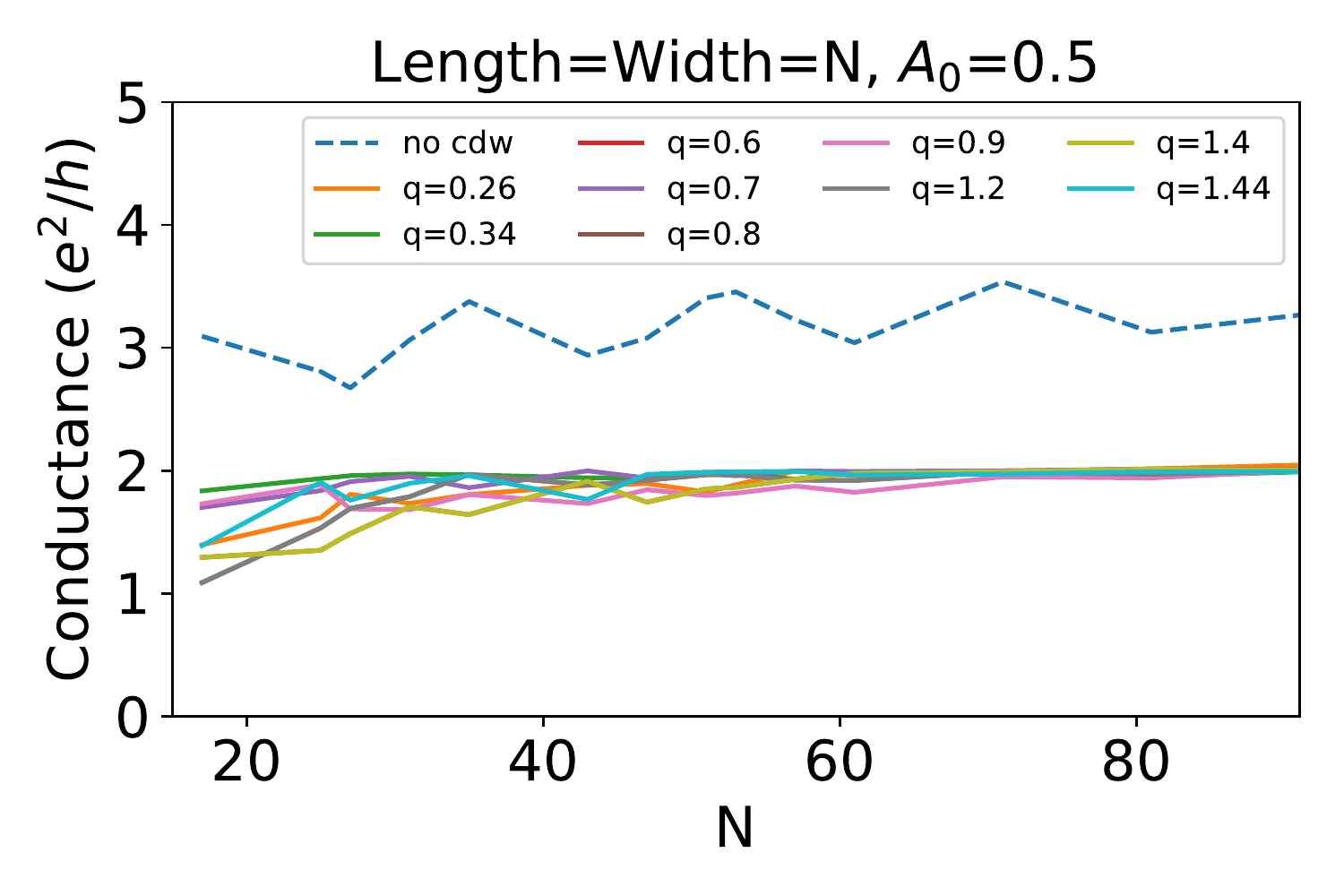}
	\caption{Conductance at energy very close to $\omega/2$ as a function of system size for different values of charge-density wave perturbation. This perturbation suppresses the contribution of bulk gapless modes and thus only the quantized contribution from gapless edge modes survive. \label{cdw}}
\end{figure}

We study the effect of a charge-density wave type perturbation in the bilayer graphene model considered above. For each site $\textbf{R}=n\textbf{a}_1+p\textbf{a}_2$ on the underlying triangular lattice of Fig.~\ref{BLG_floquetzoneedge}, we add a term of the form of $A_0\cos(qp)$ to all the nearest-neighbor hopping in the static and the drive part. We calculate the conductance in $y$ direction for different amplitudes $A_0$ and wavevector $q$ of this extra term using Kwant~\cite{groth2014kwant}. Without this charge-density wave term, the observed conductance has a contribution from both edge and bulk modes at quasienergy $\omega/2$, but  this term suppresses the bulk contribution as shown in Fig.~\ref{cdw}. In the presence of the charge-density wave perturbation, the conductance is quantized to two which indicates that it arises from the protected edge modes in the presence of a gapped bulk.

\end{document}